\documentclass[preprint]{aastex}
\shorttitle{Mm-wave line surveys toward the CND and Sgr A$^*$}
\shortauthors{Takekawa et al.}
\begin{document}
\title{Millimeter-wave Spectral Line Surveys toward the Galactic Circumnuclear Disk and Sgr A$^*$}


\author{Shunya Takekawa$^1$, Tomoharu Oka$^{1,2}$, Kunihiko Tanaka$^2$, Shinji Matsumura$^1$, Kodai Miura$^1$, and Daisuke Sakai$^3$}
\affil{
$^1$School of Fundamental Science and Technology, Graduate School of Science and Technology, Keio University, 3-14-1 Hiyoshi, Yokohama, Kanagawa 223-8522, Japan\\
$^2$Department of Physics, Institute of Science and Technology, Keio University, 3-14-1 Hiyoshi, Yokohama, Kanagawa 223-8522, Japan\\
$^3$Department of Astronomy, Graduate School of Science, The University of Tokyo, 7-3-1 Hongo, Bunkyo-ku, Tokyo 113-0033, Japan }

\email{shunya@z2.keio.jp}



\begin{abstract}
We have performed unbiased spectral line surveys at 3 mm band toward the Galactic circumnuclear disk (CND) and Sgr A$^*$ using the Nobeyama Radio Observatory (NRO) 45 m radio telescope.  The target positions are two tangential points of the CND and the direction of Sgr A$^*$.  We have obtained three wide-band spectra which cover the frequency range from 81.3 GHz to 115.8 GHz, detecting 46 molecular lines from 30 species including 10 rare isotopomers and four hydrogen recombination lines.
Each line profile consists of multiple velocity components which arise from the CND, +50 km s$^{-1}$ and +20 km s$^{-1}$ clouds (GMCs), and the foreground spiral arms.
We define the specific velocity ranges which represent the CND and the GMCs toward each direction, and classify the detected lines into three categories: the CND-/GMC-/HBD-types, based on the line intensities integrated over the defined velocity ranges.  The CND- and GMC-types are the lines which mainly trace the CND and the GMCs, respectively. The HBD-type possesses the both characteristics of the CND-/GMC-types.
We also present the lists of line intensities and other parameters, as well as intensity ratios, which must be useful to investigate the difference between nuclear environments of our Galaxy and of others.
\end{abstract}





\section{Introduction}
Most galaxies are thought to have supermassive black holes (SMBHs) at their centers, some of which are observed as active galactic nuclei (AGN).
Recent observations have provided compelling evidence that the Milky Way Galaxy also has a SMBH with mass of $(4.5\pm0.4)\times10^6 M_{\odot}$ at the dynamical center (Ghez et al. 2008; Gillessen et al. 2009), which is observed as a compact nonthermal radio source Sgr A$^*$ (Balick \& Brown 1974).
Despite the huge mass, Sgr A$^*$ is extremely dim compared to extragalactic AGN.  The luminosity of the nucleus is only $\sim10^{33-35}$ erg s$^{-1}$, which is far below the Eddington luminosity ($6\times10^{44}$ erg s$^{-1}$).  However, some authors suggested that Sgr A$^*$ had experienced highly active phases in the past.  The widespread Fe 6.4 keV fluorescent line over the central 200 pc suggests that Sgr A$^*$ was a million times brighter in X-ray than the present about several hundred years ago (Koyama et al. 1996; Murakami et al. 2000; Ryu et al. 2012). The pair of huge gamma-ray lobes recently discovered by the Fermi-LAT telescope is hypothesized to be a remnant of quasar activity at $\sim10^7$ years ago (Su et al. 2010).  It is possible that nuclear activities are transient in nature, and currently inactive Sgr A$^*$ may turn active in future.
\par Sgr A$^*$ is located in the center of the extended radio source Sgr A, which consists of a thermal “minispiral” (Sgr A West) and a nonthermal supernova remnant shell (Sgr A East).  
The circumnuclear disk (CND), which is a dense, warm ring of molecular gas, encompasses the minispiral (e.g., Genzel et al. 1985; G\"usten et al. 1987; Kaifu et al. 1987). 
The well-known 2-pc radius ring of the CND has a rotating velocity of $\sim110$ km s$^{-1}$.  The entity of the CND is possibly an infalling disk with a diameter of 10 pc, which includes the 2-pc radius ring and the negative longitude extension (Oka et al. 2011).
The CND may have been formed by the tidal capture and disruption of a molecular cloud (Sanders 1998; Wardle \& Yusef-Zadeh 2008), and is now a potential reservoir for material accreting into the central parsec.
{{It is likely that the CND is not stable but transient feature (Oka et al. 2011; Requena-Torres et al. 2012).}}
Two giant molecular clouds (GMCs), M--0.13--0.08 (+20 km s$^{-1}$ cloud) and M--0.02--0.07 (+50 km s$^{-1}$ cloud), are located in the vicinity of the nuclear region.  These GMCs are thought to be interacting with the nuclear region (e.g., G\"usten et al. 1981; Genzel et al. 1990). In particular, it is suggested that the +20 km s$^{-1}$ cloud is feeding the CND (Okumura et al. 1989; Coil \& Ho 1999).
\par  Properties of the molecular clouds in the central region are probably linked to the nuclear activities.
{{It is possible that an outflow originating from the central SMBH and the mass-losing He-stars may be interacting with the northern and southern lobes of the CND (Mu\v{z}i\'{c} et al. 2007).}}
The chemical composition of the CND provides a useful guide for research on feedback from the activity of Sgr A$^*$, as well as on the gas fueling through the CND.  High energy photons/particles generated by AGN activities form X-ray/cosmic-ray dissociation regions (XDRs/CRDRs).  Intense UV radiation forms photodissociation regions (PDRs) in irradiated molecular clouds.
Theoretical calculations predict a number of XDR/PDR discriminators, such as HCN/HCO$^+$ and HNC/HCN ratios, and abundances of NO, HOC$^+$, and HCO (Spaans \& Meijerink 2005; Meijerink et al. 2007). 
Abundance of refractory molecules, such as SiO, are thought to be enhanced in shocked regions and X-ray irradiated regions (Mart\'in et al. 2012; Amo-Baladr\'on et al. 2009).
\par Most of these diagnostic probes have transitions in millimeter wavelength.  Using some of the diagnostic probes with intense emission, several authors have studied the chemistry of the CND and adjacent GMCs (e.g., Amo-Baladr\'on et al. 2011; Mart\'in et al. 2012).  Amo-Baladr\'on et al. (2011) suggested a possible connection between the CND and the GMCs.  In order to investigate their physical connection, a careful search for the best  probes of the CND and the GMCs must be essential.  Unbiased spectral lines surveys are useful to search for the probes, and to diagnose the physical conditions and chemical composition.  However, no unbiased line surveys in mm-wave range toward the CND and Sgr A$^*$ have been reported so far.  
Therefore, we have performed unbiased spectral line surveys at 3 mm band toward the CND and Sgr A$^*$.
This paper focuses on full presentation of obtained spectra and line parameters between characteristic velocity ranges.  Detailed analyses of physical conditions and chemical composition will be presented in forthcoming papers.

\section{Observations}
The observations were carried out in February and May 2013 with the Nobeyama Radio Observatory (NRO) 45 m radio telescope. Target positions were chosen as $(\Delta l$,$\Delta b)$$=$$(+46''$,$0'')$, $(0''$,$0'')$, $(-40''$,$0'')$, which are defined by offsets from the position of Sgr A$^*$, $(\alpha_{2000}$, $\delta_{2000})$$=$$(17^{\rm h}$ $45^{\rm m}$ $40^{\rm s}.0$, $-29^\circ$ $00'$ $27''.9$) (Reid \& Brunthaler 2004). Hereafter, we refer those target positions to NE, SGA, SW, respectively (Figure 1). 

\begin{figure}[htb]
\begin{center}
\epsscale{1}
\includegraphics[width=15cm]{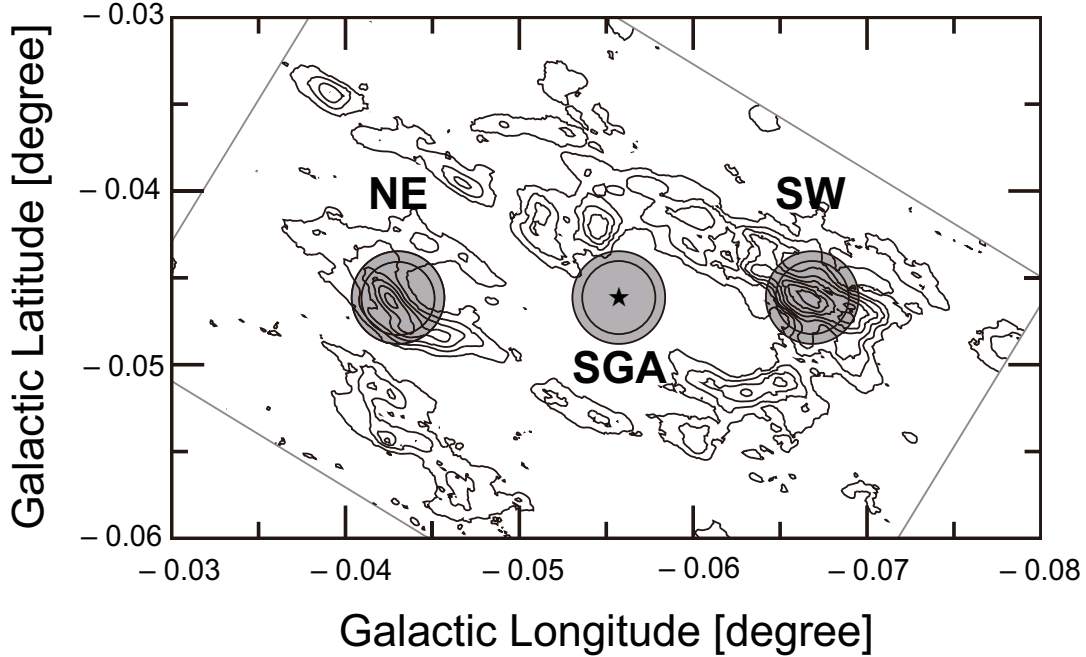}
\caption{Target positions of our line surveys (circles) superposed on a map of the velocity-integrated HCN $J = 1-0$ emission (Christopher et al. 2005). The star indicates the position of Sgr A$^*$. Sizes of circles represent the half-power beam widths (HPBW) of the NRO 45 m radio telescope at 115 GHz and 86 GHz.}
\end{center}
\end{figure}

\par For these line surveys, we used the TZ1 V/H receivers which were operated in the two-side band mode. The beamwidth and the main-beam efficiency ($\eta_{\rm MB}$)  are listed in Table 1. We used the SAM45 spectrometer in the 2 GHz bandwidth (488.24 kHz resolution) mode. Using 14 arrays of SAM45, we obtained a 4 GHz instantaneous bandwidth for each sideband (LSB and USB).  
The system noise temperatures ranged from 120 to 360 K during the observations.  The observations were made by the 3:1 on-off position switching mode.  The reference positions were $(l, b)=(0^\circ.0, +0^\circ.5), (0^\circ.0, -0^\circ.5)$, which were observed alternately.  Pointing errors were corrected every 1.5 hours by observing the SiO maser emissions (43 GHz) from VX Sgr with the H40 receiver. The pointing accuracy was better than 3$''$ (rms) in both azimuth and elevation. Calibration of the antenna temperature was accomplished by the standard chopper-wheel method.

\begin{deluxetable}{cccc}
\tablecolumns{10}
\tablewidth{0pc}
\tablecaption{Beamwidth and main beam efficiency of the 45-m RT}
\tablehead{ \colhead{Receiver}  & \colhead{Frequency} & \colhead{Beamwidth}  & \colhead{$\eta_{\rm MB}$}\\
\colhead{} & \colhead{(GHz)} & \colhead{(arcsec)}  & \colhead{(\%)}}
\startdata
 & 86 & 19.2$\pm0.2$ & 36$\pm$2\\
TZ1(H) &110&16.1$\pm$0.1&31$\pm$2\\
 &115&15.4$\pm$0.1&28$\pm$1\\
\tableline
 &86&19.2$\pm0.2$&36$\pm$2\\
TZ1(V) &110&16.3$\pm$0.1&32$\pm$3\\
 &115&14.9$\pm$0.1&29$\pm$2
\enddata
\end{deluxetable}

\par All data were reduced using the NEWSTAR reduction package. We subtracted baselines of each 2 GHz spectrum by fitting sixth- or seventh-degree polynomials.  We composed wide-band spectra covering 81.3--115.8 GHz toward the observed three positions by averaging and merging all 2 GHz spectra (Figure 2). The spectral resolution of the resultant spectra was 1.0 MHz.  Antenna temperatures ($T^*_{a}$) were converted into main-beam temperatures ($T_{\rm MB}$) by multiplying $1/\eta_{\rm MB}(\nu)$.
For the frequency dependence of the efficiency $\eta_{\rm MB}$, we adopted that $\eta_{\rm MB}(\nu)$ $=$ $-0.236 \times (\nu_{\rm obs}/\rm GHz) + 56.47$ (\%), which was obtained by the least-square fitting to the $\eta_{\rm MB}$ measured at three frequencies (Table 1). The rms noise levels of the wide-band spectra were calculated by using data in emission/absorption-free frequency ranges (Figure 3).

\begin{figure*}[h]
\begin{center}
\vspace{1.5cm}
\rotatebox{90}{
\centering{
\begin{minipage}{20cm}
\epsscale{1}
\includegraphics[width=20cm]{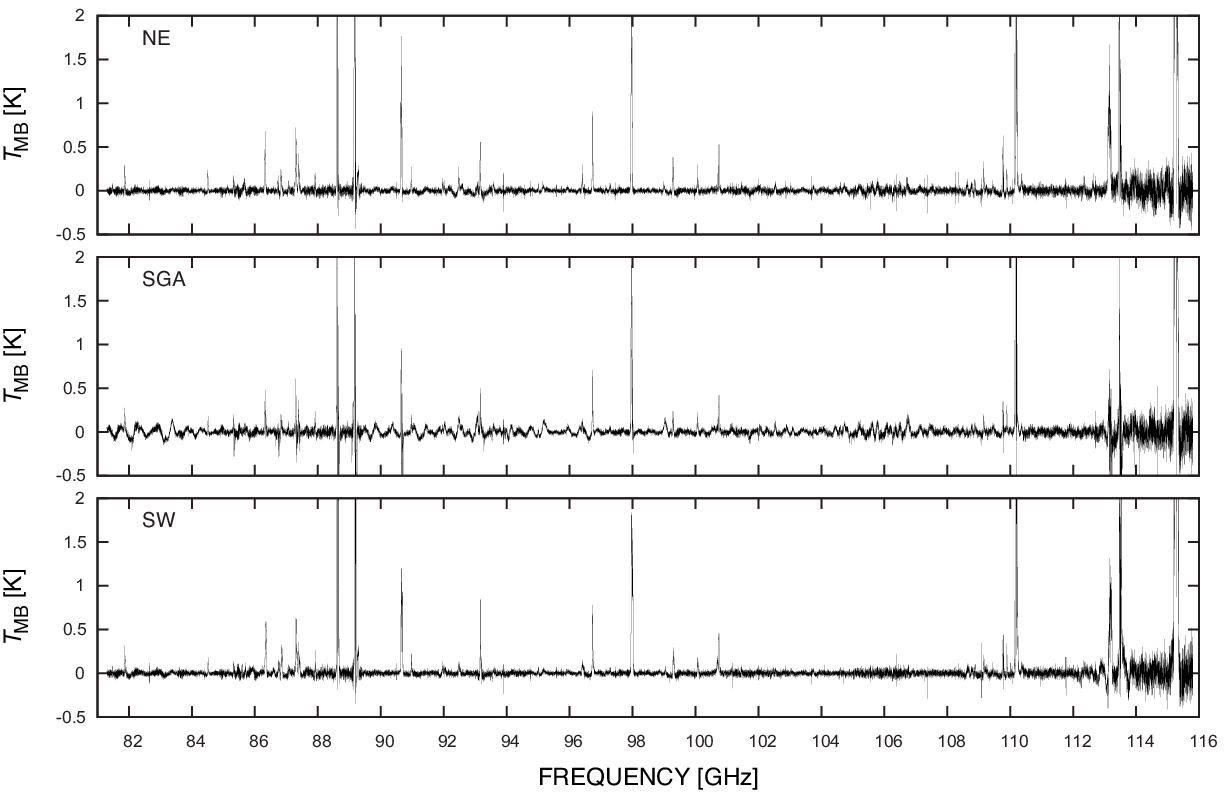}
\caption{Spectral line zoos obtained at three observed positions, NE, SW and SGA.}
\end{minipage}
}}
\end{center}
\end{figure*}

\begin{figure}[htb]
\begin{center}
\epsscale{1}
\includegraphics{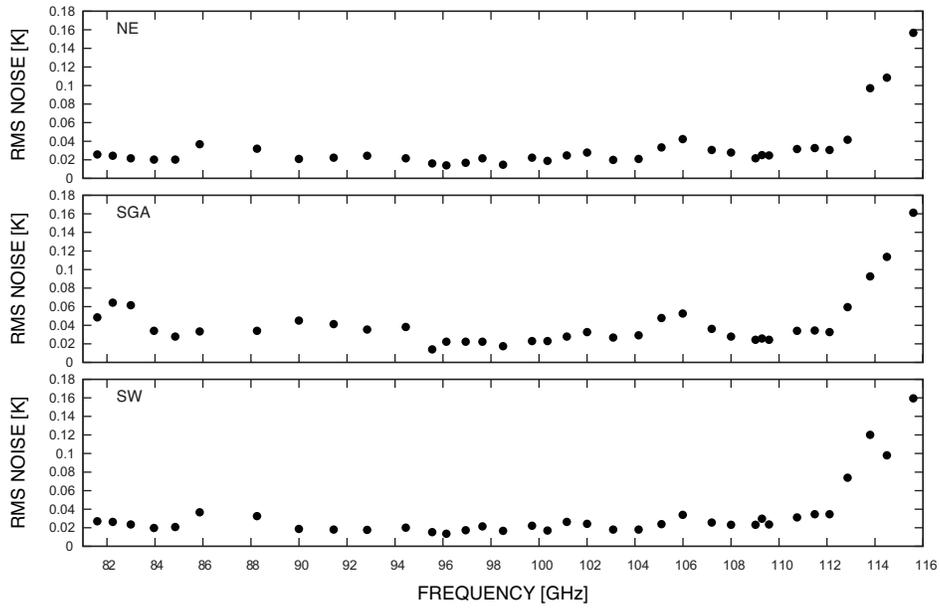}
\caption{Plots of rms noise of $T_{\rm MB}$ at three observed positions. Each value is calculated in relatively stable baseline without spectral lines.}
\end{center}
\end{figure}

\clearpage
\section{Results}
We show the wide-band spectra toward NE, SW, and SGA in Figure 2.
A number of spectral lines appear in these wide-band spectra.  The SGA spectrum is more rippled than the other positions, presumably because the intense continuum emission from Sgr A$^*$ manifests the non-linearity of the system.  We identified four hydrogen recombination lines and 46 lines from 30 molecular species including 10 rare isotopomers (Table 2).  In addition to familiar diatomic and triatomic molecules, {{a few more complex molecules}} (e.g., HCCCN, c-C$_3$H$_2$, CH$_3$OH, $^{13}$CH$_3$OH, CH$_3$CHO, and CH$_3$CN) were detected.

\par We extracted 50 line profiles for each position (Figure 4). Each extracted spectrum covers the LSR velocity range from $-$300 to $+$300 km s$^{-1}$. After the extraction, we again subtracted baselines from the spectra by fitting up to third-degree polynomials, although a linear function was used in most cases. 
{{Flat lines in some extracted spectra denote velocity range where adjacent lines appear.}}
The peak velocities ($V_{\rm peak}$), velocity dispersions ($\sigma_{V}$), peak temperatures ($T_{\rm peak}$), and 1$\sigma$ rms noise levels ($\Delta T_{\rm MB}$) of the detected lines are listed in Tables 2, 3.  The $\Delta T_{\rm MB}$ values were calculated for the velocity ranges from $-$300 to $-$200 km s$^{-1}$ and from $+$200 to $+$300 km s$^{-1}$. The $\sigma_{V}$ values were calculated using pixels with $T_{\rm MB} > 3\Delta T_{\rm MB}$.
All the detected lines have larger velocity widths than those from typical molecular clouds in the Galactic disk. Because of their large velocity widths, some lines are blended and/or unresolved.
The lines of common interstellar molecules (e.g., CO, CS, HCN, and HCO$^+$) are intense ($T_{\rm peak} \gtrsim1$ K) and have very large velocity widths at every position.
In particular, CO $J$=1--0 line exceeds 35 K and shows a number of velocity components.  About a half of the lines from SGA show absorption features due to the foreground gas in the Galactic disk.

\par Overall velocity structures of the line profiles vary mainly depending on the position. The profiles of the detected lines, except for the hydrogen recombination lines, contain similar velocity components in each position. Most molecular lines from NE peak at $V_{\rm LSR}\sim$+50 km s$^{-1}$, while those from SW peak at $V_{\rm LSR}\sim$+20 km s$^{-1}$. Molecular lines from SGA generally exhibit complex profiles.   The hydrogen recombination lines toward SGA show very large velocity width, which may be attributed to the rapidly rotating minispiral.  The biased velocities of the NE and SW recombination lines, which peak at $V_{\rm LSR}$$\sim$+80 and $\sim-80$ km s$^{-1}$, respectively (except for H41$\alpha$ from SW), may suggest that rotating ionized gas is associated with the CND.

\begin{figure*}[c]
\begin{center}
\epsscale{1}
\includegraphics[height=20cm]{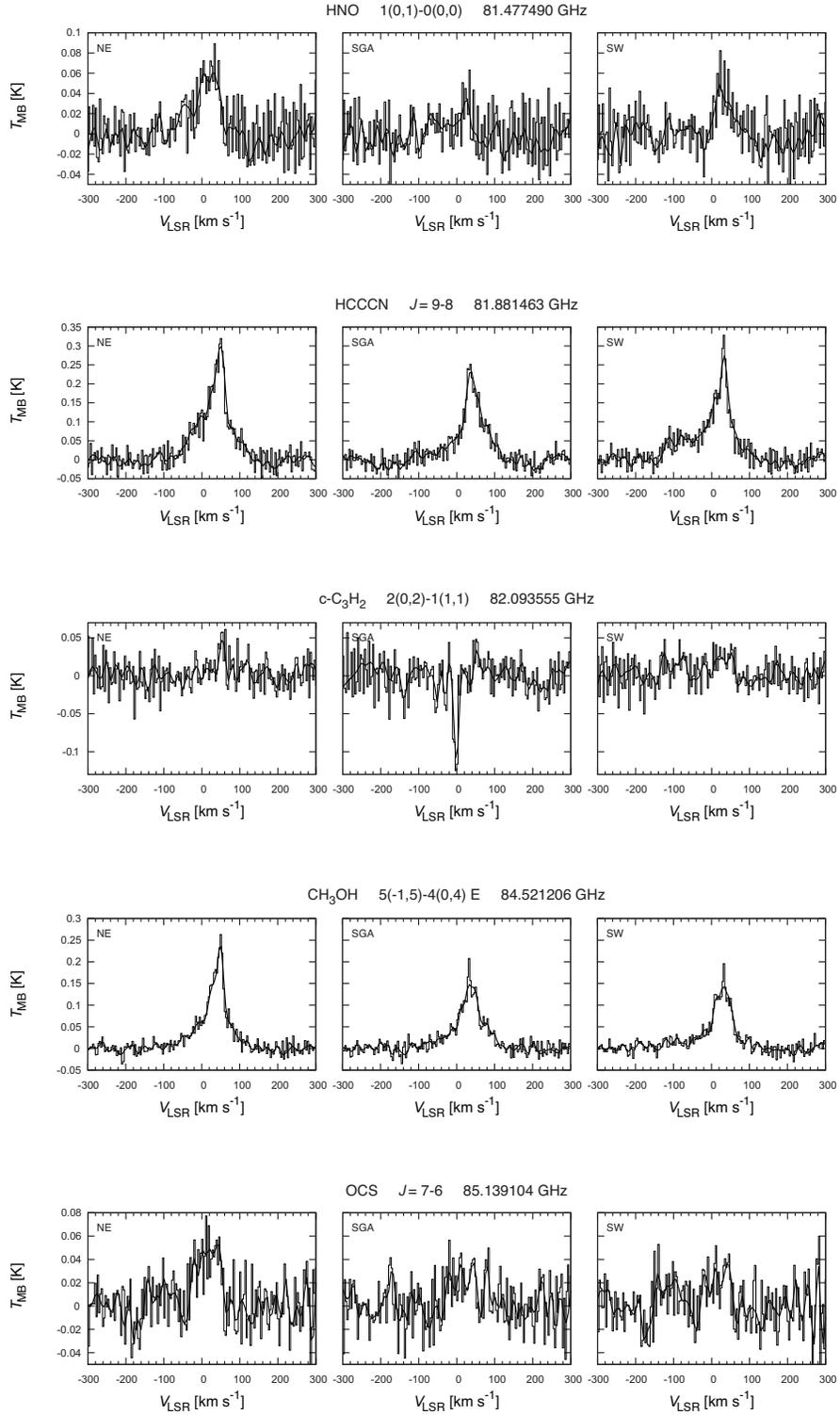}
\caption{Spectral line profiles toward three observed positions. Thick lines are smoothed spectra with grid width of 10 km s$^{-1}$.
{Flat lines in some extracted spectra denote velocity ranges where adjacent lines appear.}}
\end{center}
\end{figure*}

\setcounter{figure}{3}
\begin{figure*}[c]
\begin{center}
\epsscale{1}
\includegraphics[height=20cm]{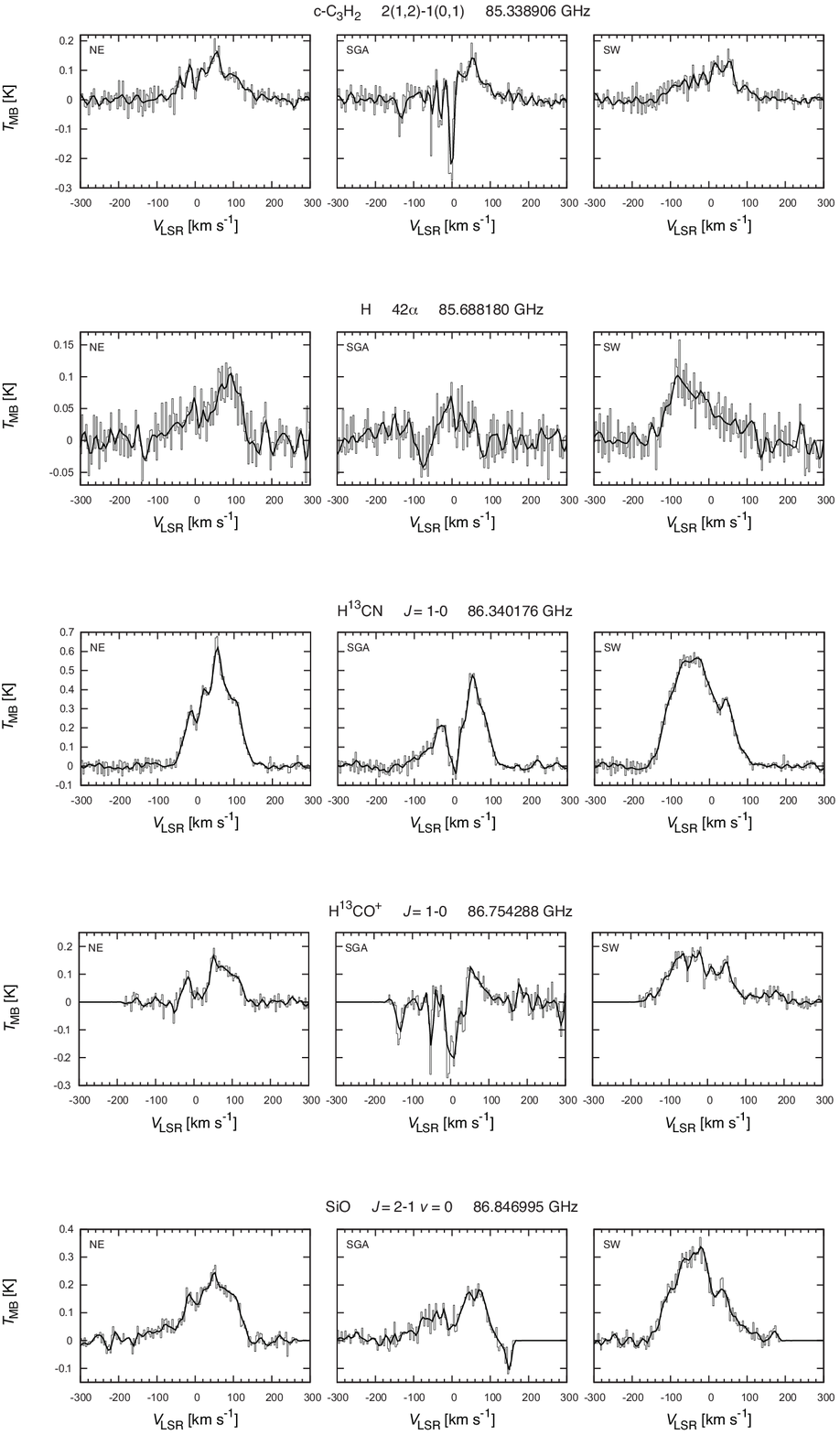}
\caption{Continued}
\end{center}
\end{figure*}

\setcounter{figure}{3}
\begin{figure*}[c]
\begin{center}
\epsscale{1}
\includegraphics[height=20cm]{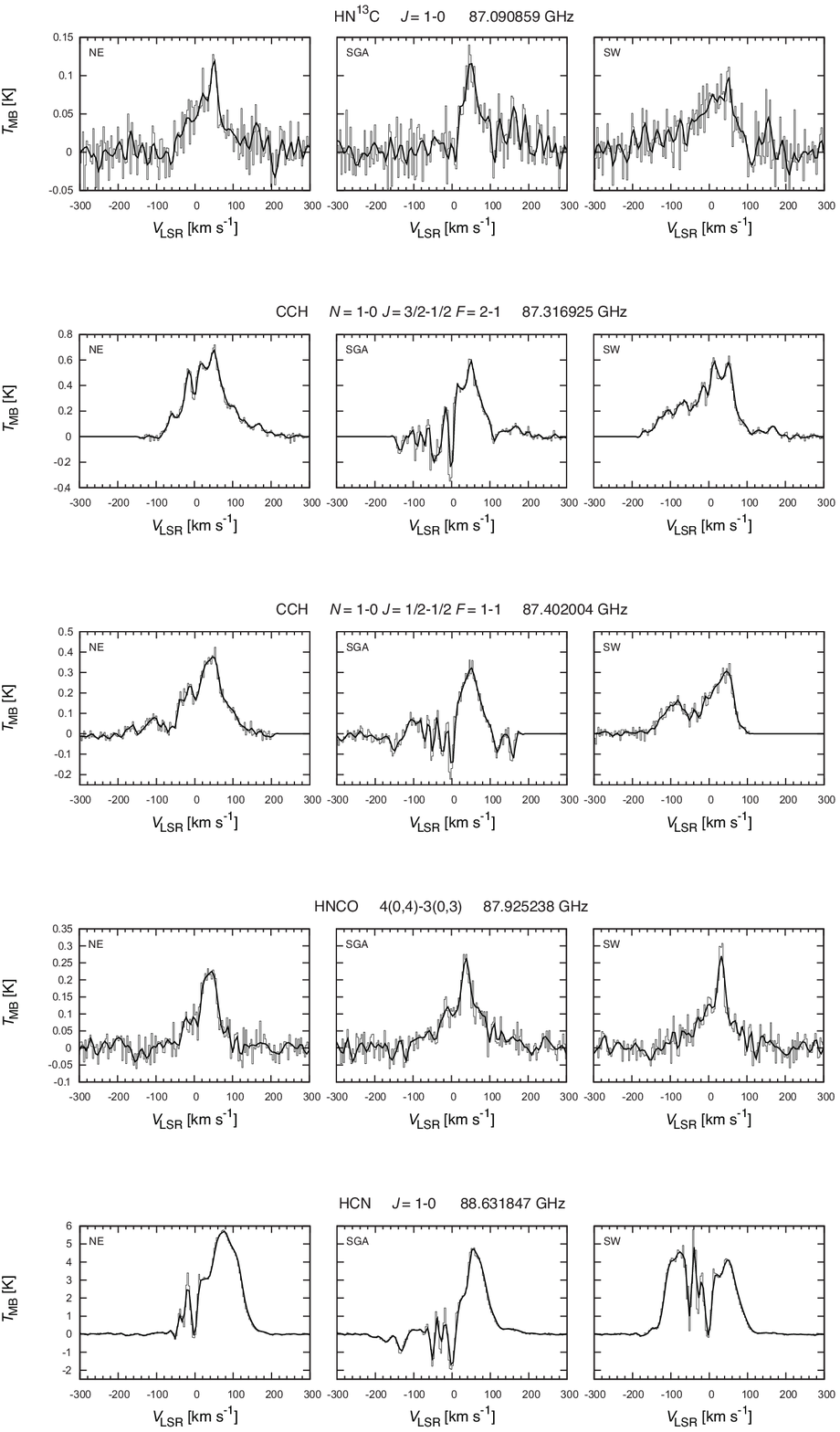}
\caption{Continued}
\end{center}
\end{figure*}

\setcounter{figure}{3}
\begin{figure*}[c]
\begin{center}
\epsscale{1}
\includegraphics[height=20cm]{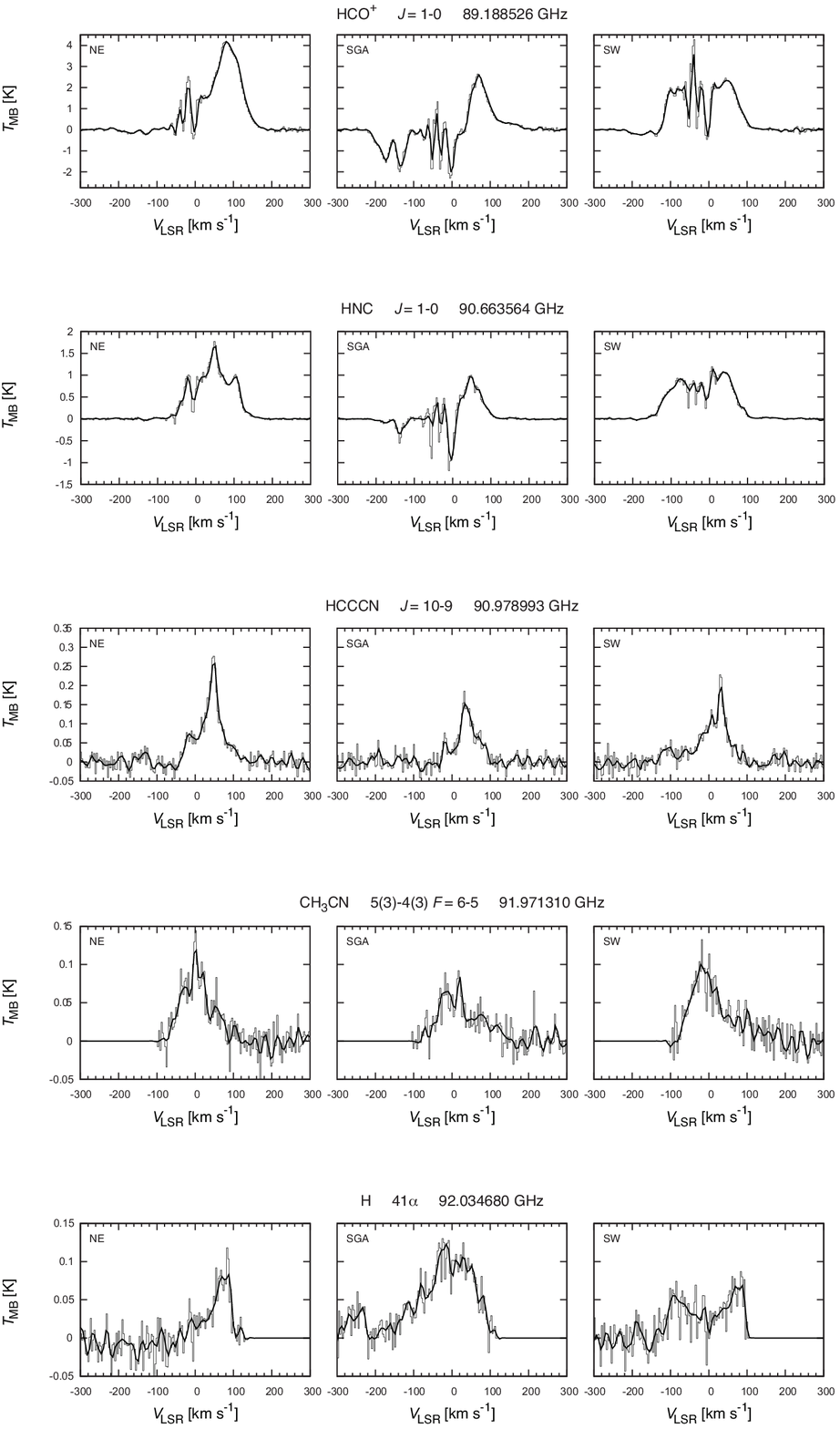}
\caption{Continued}
\end{center}
\end{figure*}

\setcounter{figure}{3}
\begin{figure*}[c]
\begin{center}
\epsscale{1}
\includegraphics[height=20cm]{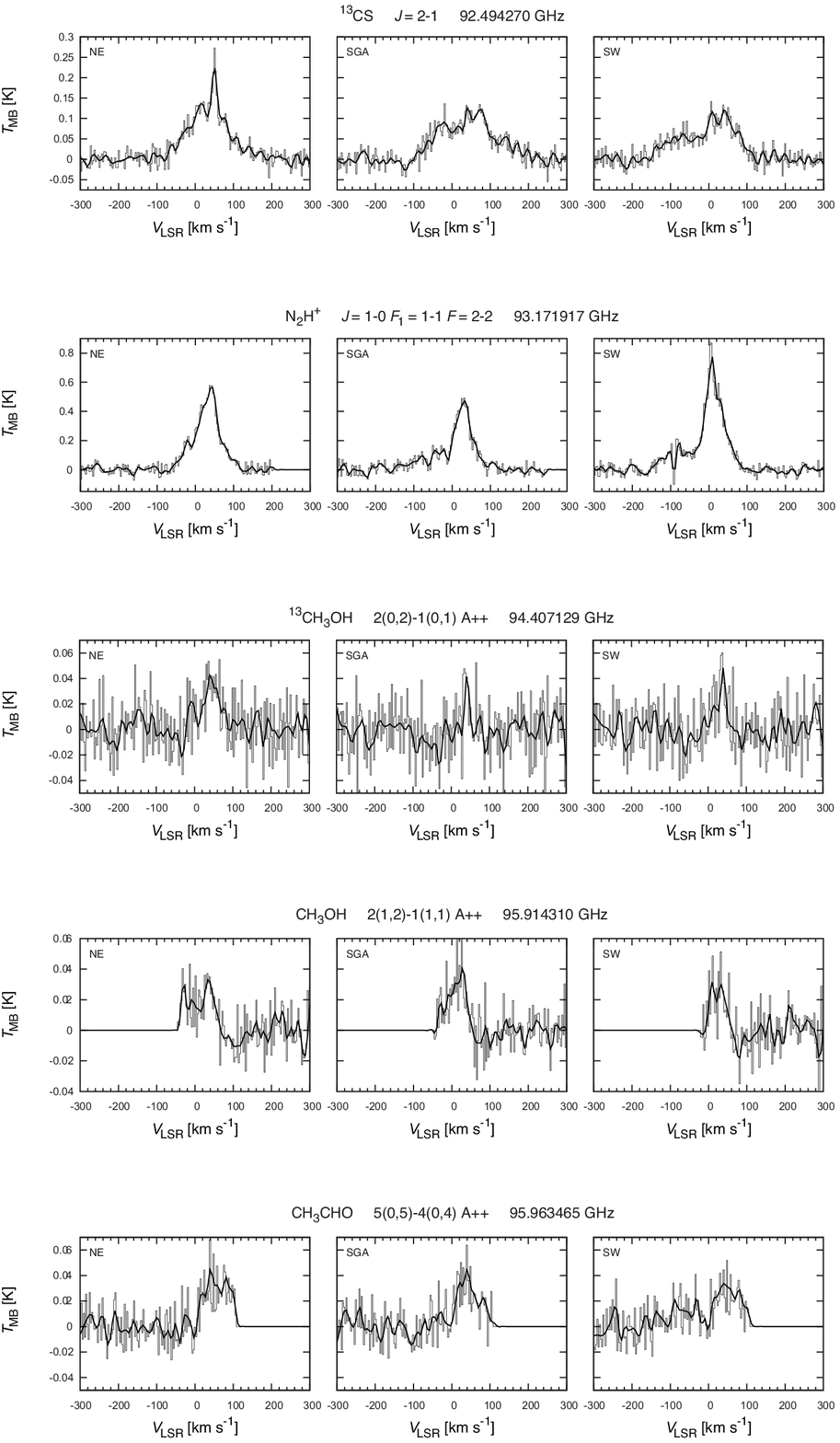}
\caption{Continued}
\end{center}
\end{figure*}

\setcounter{figure}{3}
\begin{figure*}[c]
\begin{center}
\epsscale{1}
\includegraphics[height=20cm]{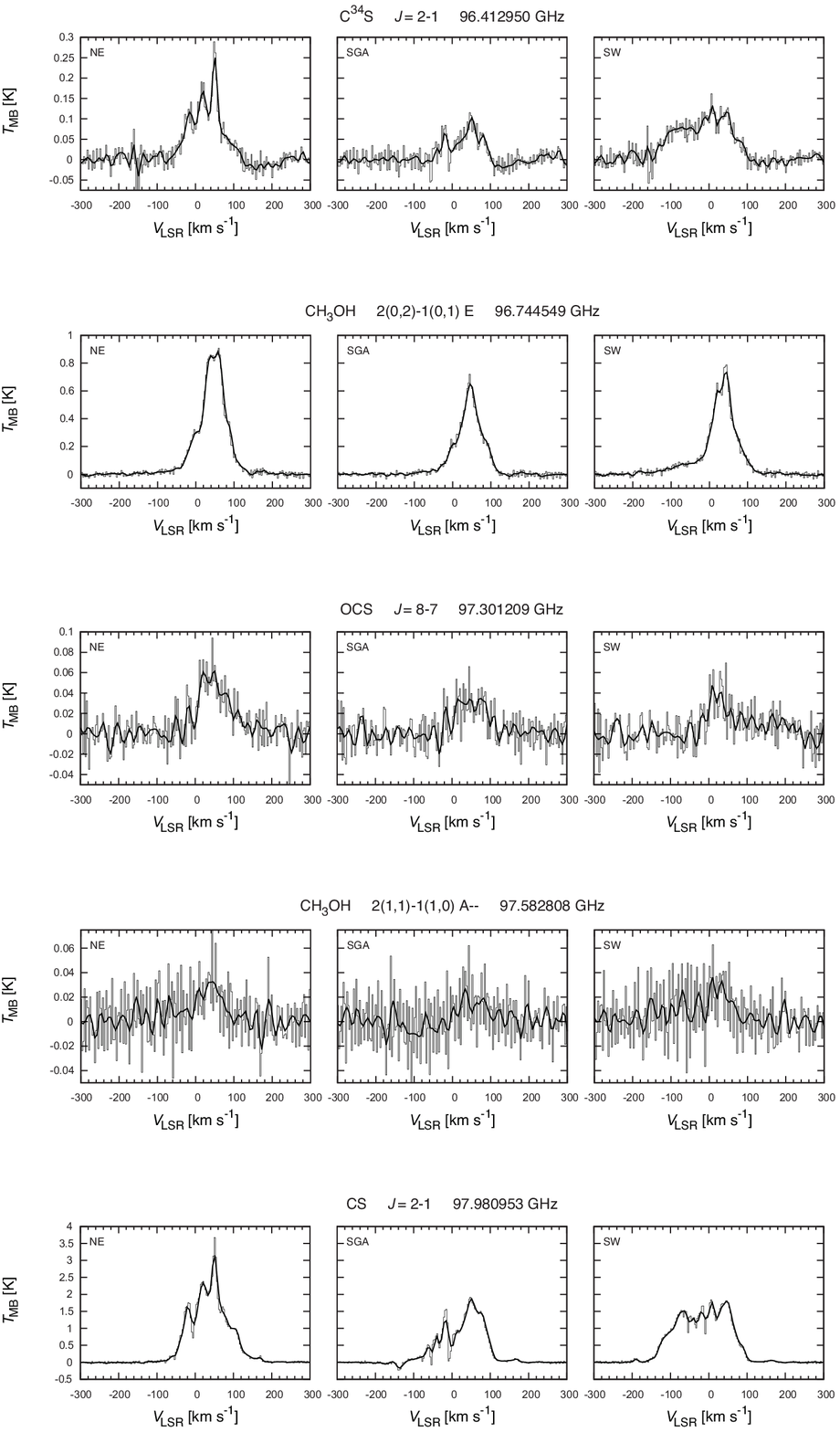}
\caption{Continued}
\end{center}
\end{figure*}

\setcounter{figure}{3}
\begin{figure*}[c]
\begin{center}
\epsscale{1}
\includegraphics[height=20cm]{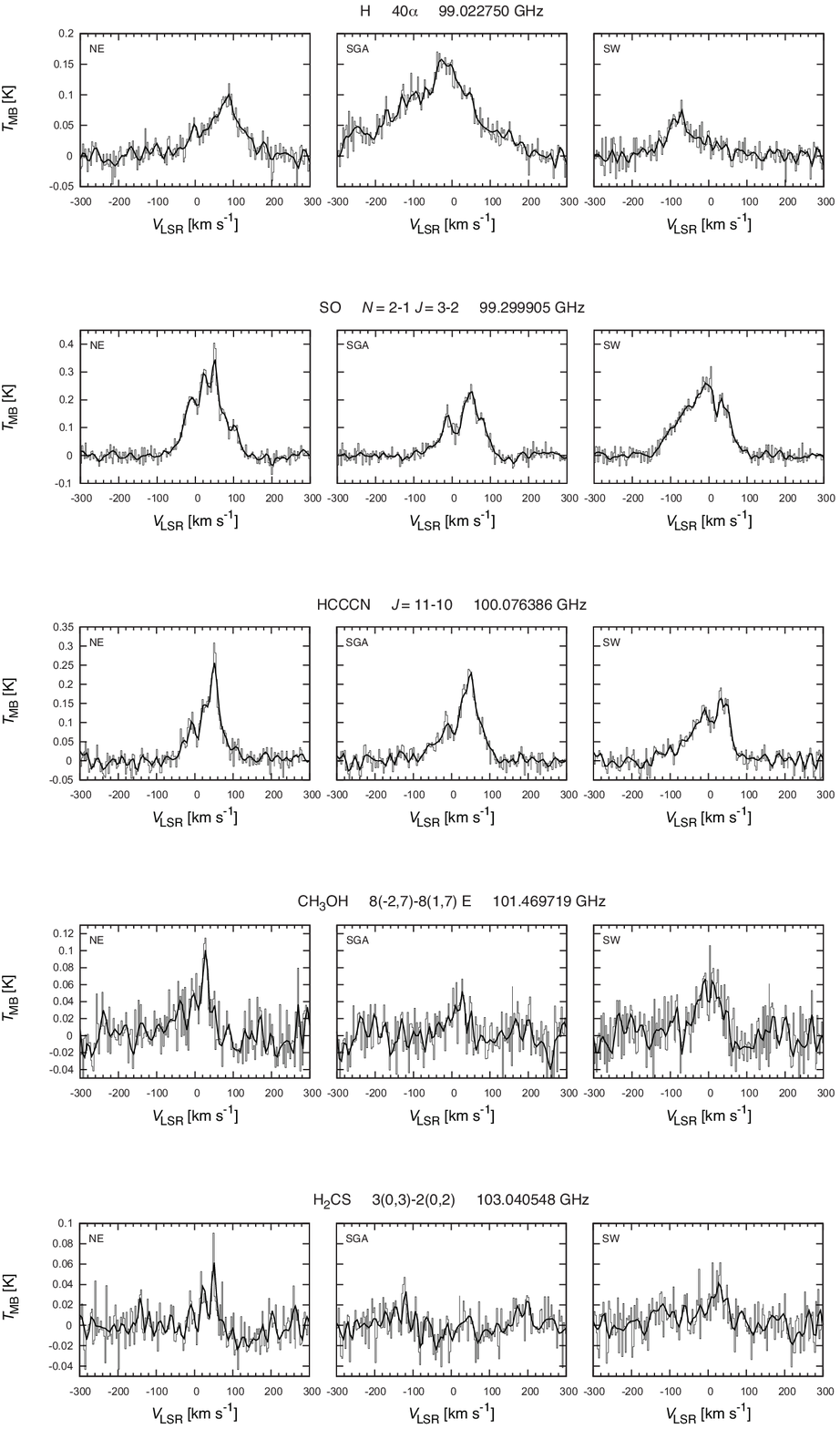}
\caption{Continued}
\end{center}
\end{figure*}

\setcounter{figure}{3}
\begin{figure*}[c]
\begin{center}
\epsscale{1}
\includegraphics[height=20cm]{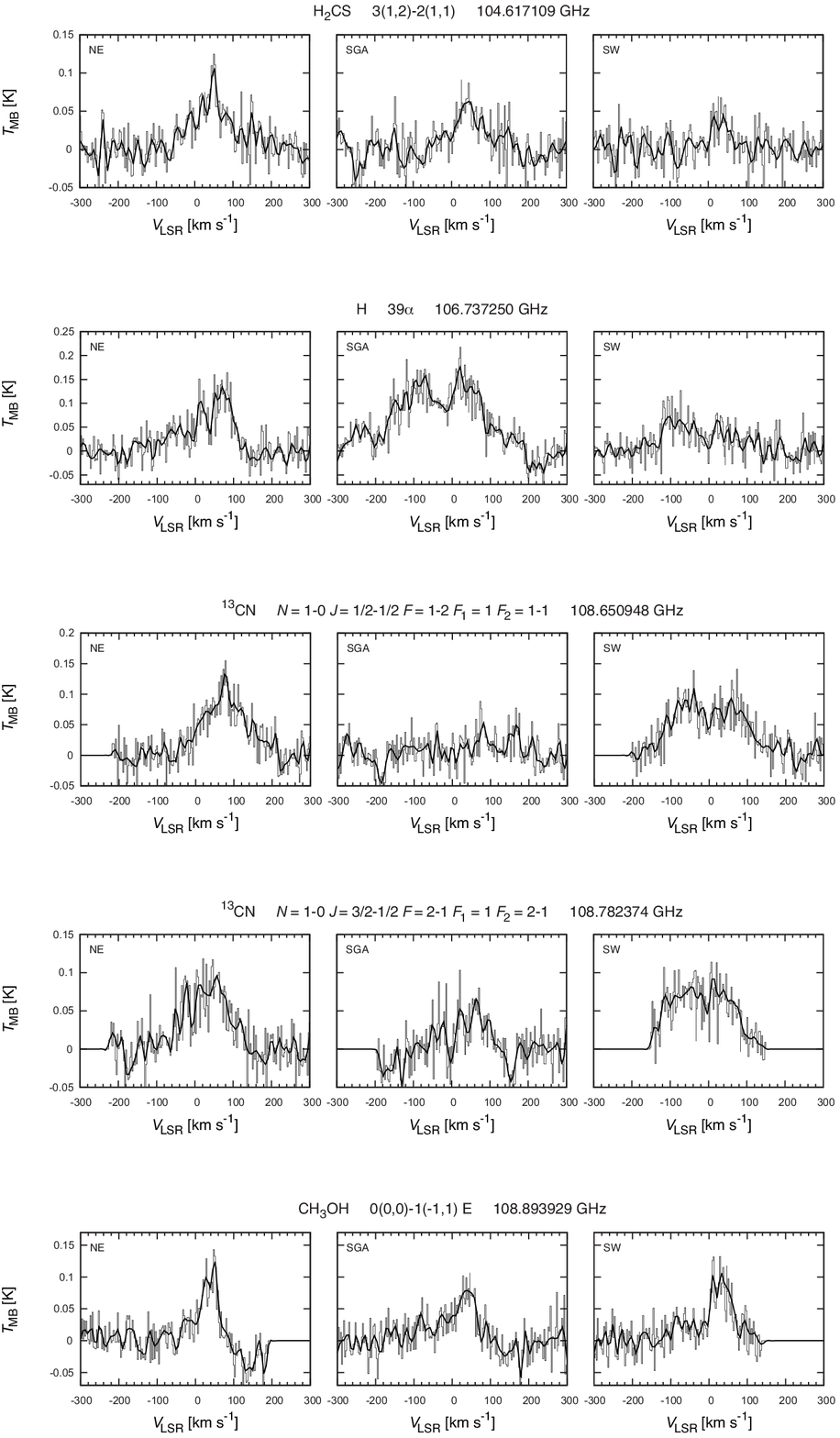}
\caption{Continued}
\end{center}
\end{figure*}

\setcounter{figure}{3}
\begin{figure*}[c]
\begin{center}
\epsscale{1}
\includegraphics[height=20cm]{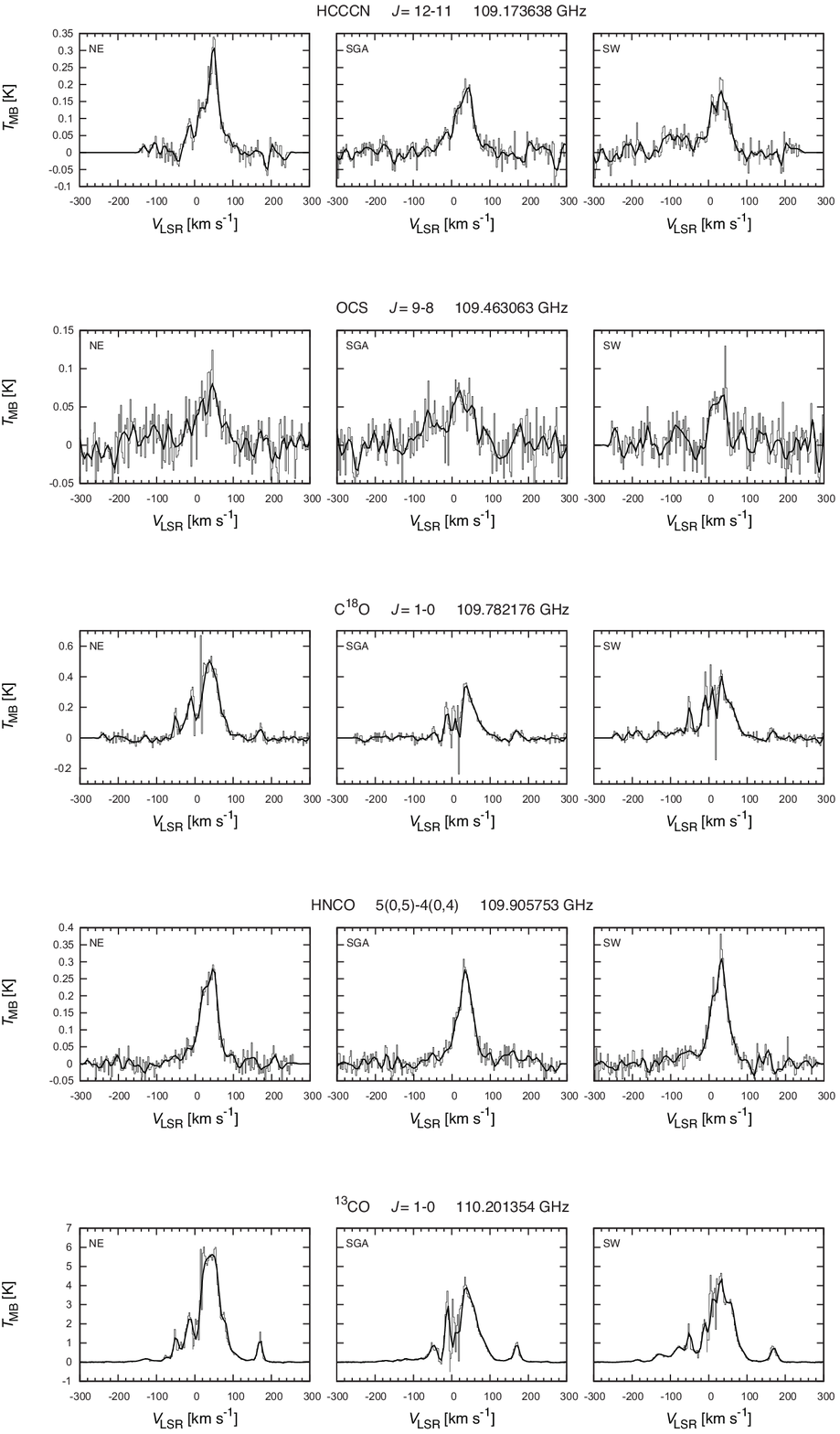}
\caption{Continued}
\end{center}
\end{figure*}

\setcounter{figure}{3}
\begin{figure*}[c]
\begin{center}
\epsscale{1}
\includegraphics[height=20cm]{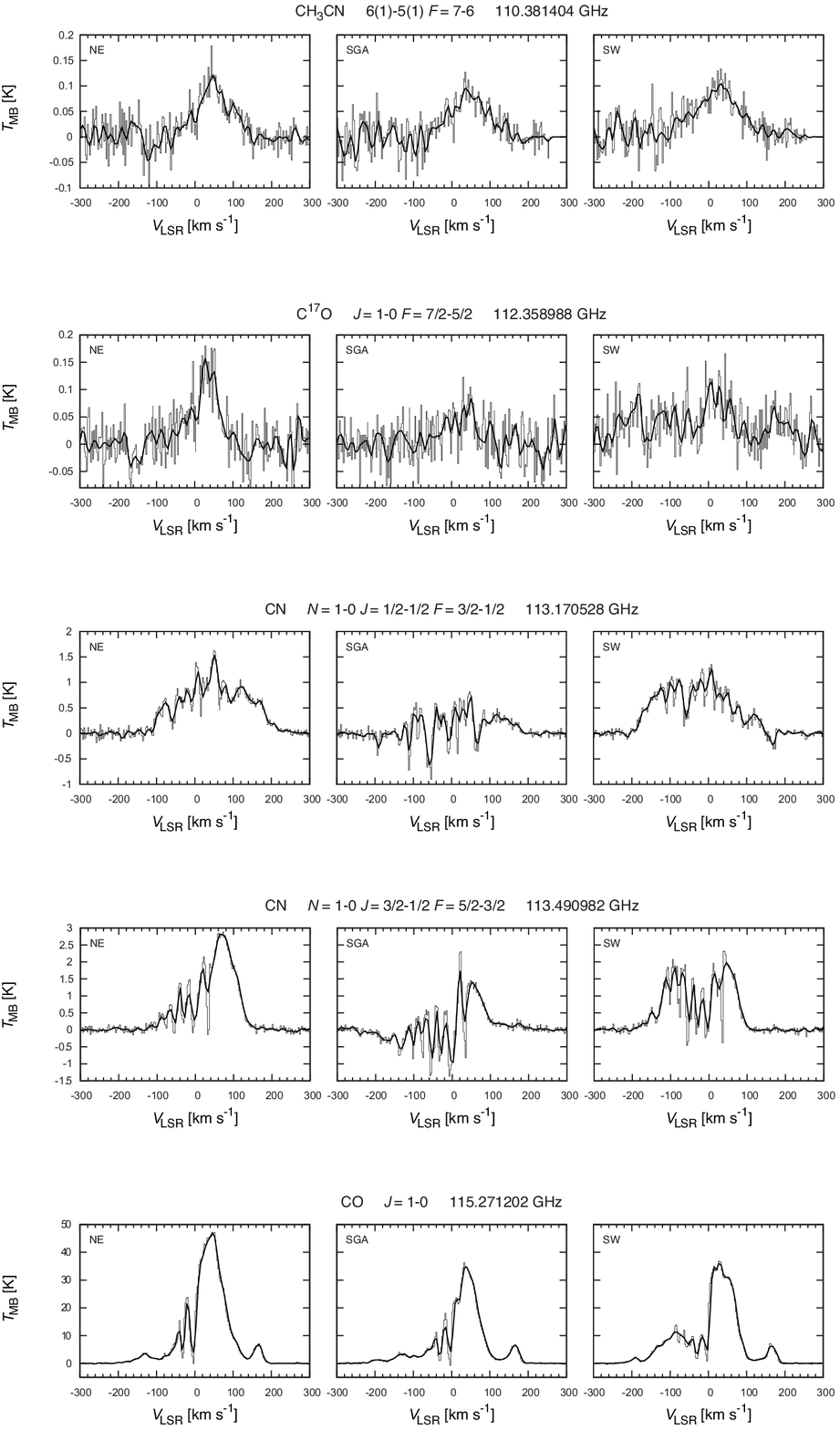}
\caption{Continued}
\end{center}
\end{figure*}

\clearpage
\begin{deluxetable}{rccccccccc}
\tablecolumns{10}
\tablewidth{0pc}
\tablecaption{Peak velocities and velocity dispersions of the identified spectral lines}
\tablehead{
\colhead{} &  \colhead{}	& \colhead{Rest Freq.}	&	\multicolumn{3}{c}{$V_{\rm peak}$ (km s$^{-1}$)} &\colhead{} & \multicolumn{3}{c}{$\sigma_{V}$ (km s$^{-1}$)} \\
\cline{4-6} \cline{8-10} 																	
\colhead{Species}	&	\colhead{Transition}	&	\colhead{(GHz)}	&	\colhead{NE}	&	\colhead{SGA}	&	\colhead{SW}	&\colhead{}&	\colhead{NE}	&	\colhead{SGA}	& \colhead{SW}	
}
\startdata																	
HNO	&	$1_{0,1}-0_{0,0}$	&	81.477490	&	35.65	&	35.65	&	24.98	&&	--	&	--	&	--	\\
HCCCN	&	$J=9-8$	&	81.881463	&	52.22	&	37.94	&	34.28	&&	31.37	&	25.13	&	41.16	\\
c-C$_{3}$H$_{2}$	&	$2_{0,2}-1_{1,1}$	&	82.093555	&	63.74	&	52.79	&	-82.70	&&	--	&	--	&	--	\\
CH$_3$OH	&	$5_{-1,5}-4_{0,4}$E	&	84.521206	&	52.16	&	34.78	&	34.78	&&	24.70	&	27.42	&	19.58	\\
OCS	&	$J=7-6$	&	85.139104	&	13.39	&	-17.59	&	13.39	&&	--	&	--	&	--	\\
c-C$_3$H$_2$	&	$2_{1,2}-1_{0,1}$	&	85.338906	&	53.07	&	53.07	&	53.07	&&	57.70	&	28.03	&	55.83	\\
H	&	$42\alpha$	&	85.688180	&	82.50	&	0.63	&	-74.59	&&	31.44	&	--	&	32.31	\\
H$^{13}$CN	&	$J=1-0$	&	86.340176	&	57.21	&	57.21	&	-37.58	&&	42.37	&	49.27	&	54.49	\\
H$^{13}$CO$^{+}$	&	$J=1-0$	&	86.754288	&	54.21	&	47.30	&	-16.63	&&	40.51	&	--	&	62.09	\\
SiO	&	$J=2-1$	&	86.846995	&	53.83	&	70.75	&	-20.38	&&	48.00	&	57.09	&	52.24	\\
HN$^{13}$C	&	$J=1-0$	&	87.090859	&	49.43	&	45.99	&	52.87	&&	21.47	&	39.47	&	45.28	\\
CCH	&	$N_{J,F}=1_{3/2,2}-0_{1/2,1}$	&	87.316925	&	54.33	&	47.81	&	54.33	&&	51.80	&	39.52	&	63.38	\\
CCH	&	$N_{J.F}=1_{1/2,1}-0_{1/2,1}$	&	87.402004	&	54.89	&	48.03	&	54.89	&&	55.99	&	42.15	&	60.36	\\
HNCO	&	$4_{0,4}-3_{0,3}$	&	87.925238	&	37.29	&	34.23	&	37.29	&&	33.81	&	34.60	&	32.56	\\
HCN	&	$J=1-0$	&	88.631847	&	75.59	&	59.01	&	-40.09	&&	44.96	&	44.17	&	65.52	\\
HCO$^+$	&	$J=1-0$	&	89.188526	&	82.10	&	68.99	&	-36.22	&&	47.24	&	39.32	&	60.43	\\
HNC	&	$J=1-0$	&	90.663564	&	50.47	&	50.47	&	8.48	&&	45.39	&	41.53	&	60.97	\\
HCCCN	&	$J=10-9$	&	90.978993	&	50.39	&	34.25	&	30.95	&&	25.35	&	42.54	&	26.91	\\
CH$_3$CN	&	$J_{K,F}=5_{3,6}-4_{3,5}$	&	91.971310	&	2.97	&	18.94	&	-16.27	&&	25.01	&	25.87	&	32.79	\\
H	&	$41\alpha$	&	92.034680	&	85.28	&	-22.87	&	85.28	&&	11.17	&	37.96	&	71.98	\\
$^{13}$CS	&	$J=2-1$	&	92.494270	&	52.09	&	-17.28	&	8.01	&&	36.52	&	52.87	&	52.29	\\
N$_2$H$^{+}$	&	$J_{F_1,F}=1_{1,2}-0_{1,2}$	&	93.171917	&	39.31	&	32.87	&	7.78	&&	29.82	&	29.79	&	39.85	\\
$^{13}$CH$_3$OH	&	$2_{0,2}-1_{0,1}$A++	&	94.407129	&	66.78	&	66.78	&	38.83	&&	--	&	--	&	--	\\
CH$_3$OH	&	$2_{1,2}-1_{1,1}$A++	&	95.914310	&	-12.16	&	15.35	&	33.79	&&	--	&	--	&	--	\\
CH$_3$CHO	&	$5_{0,5}-4_{0,4}$A++	&	95.963465	&	40.82	&	40.82	&	49.88	&&	--	&	--	&	--	\\
C$^{34}$S	&	$J=2-1$	&	96.412950	&	50.84	&	47.73	&	8.24	&&	38.64	&	32.53	&	56.44	\\
CH$_3$OH	&	$2_{0,2}-1_{0,1}$E	&	96.744549	&	61.51	&	46.32	&	46.32	&&	32.96	&	31.11	&	42.92	\\
OCS	&	$J=8-7$	&	97.301209	&	46.24	&	46.24	&	46.24	&&	22.21	&	--	&	15.04	\\
CH$_3$OH	&	$2_{1,1}-1_{1,0}$A$--$	&	97.582808	&	44.57	&	44.57	&	11.39	&&	--	&	--	&	47.27	\\
CS	&	$J=2-1$	&	97.980953	&	52.48	&	49.42	&	7.81	&&	45.37	&	49.44	&	60.44	\\
H	&	$40\alpha$	&	99.022750	&	90.07	&	-37.09	&	-69.48	&&	29.24	&	58.19	&	22.41	\\
SO	&	$N_J=2_3-1_2$	&	99.299905	&	51.04	&	51.04	&	6.96	&&	35.88	&	34.19	&	45.19	\\
HCCCN	&	$J=11-10$	&	100.076386	&	52.08	&	46.09	&	34.11	&&	26.50	&	29.91	&	29.07	\\
CH$_3$OH	&	$8_{-2,7}-8_{1,7}$E	&	101.469719	&	28.71	&	28.71	&	5.08	&&	--	&	--	&	--	\\
H$_2$CS	&	$3_{0,3}-2_{0,2}$	&	103.040548	&	51.06	&	21.96	&	33.60	&&	--	&	--	&	--	\\
H$_2$CS	&	$3_{1,2}-2_{1,1}$	&	104.617109	&	51.89	&	26.10	&	26.10	&&	40.34	&	--	&	--	\\
H	&	$39\alpha$	&	106.737250	&	84.96	&	23.17	&	-72.32	&&	37.68	&	68.65	&	50.55	\\
$^{13}$CN	&	$N_{J,F}=1_{1/2,1}-0_{1/2,2}$	&	108.658948	&	79.87	&	74.35	&	74.35	&&	32.88	&	--	&	60.37	\\
$^{13}$CN	&	$N_{J,F}=1_{3/2,2}-0_{1/2,1}$	&	108.782374	&	23.08	&	23.08	&	9.30	&&	42.86	&	39.26	&	--	\\
CH$_3$OH	&	$0_{0,0}-1_{-1,1}$E	&	108.893929	&	49.36	&	46.61	&	30.09	&&	13.98	&	--	&	15.89	\\
HCCCN	&	$J=12-11$	&	109.173638	&	51.18	&	37.45	&	31.96	&&	22.29	&	14.04	&	47.41	\\
OCS	&	$J=9-8$	&	109.463063	&	46.73	&	46.73	&	43.99	&&	19.87	&	34.31	&	--	\\
C$^{18}$O	&	$J=1-0$	&	109.782176	&	16.87	&	38.71	&	5.94	&&	35.23	&	28.78	&	35.12	\\
HNCO	&	$5_{0,5}-4_{0,4}$	&	109.905753	&	48.43	&	32.06	&	32.06	&&	22.70	&	18.19	&	17.77	\\
$^{13}$CO	&	$J=1-0$	&	110.201354	&	25.45	&	36.33	&	33.61	&&	49.48	&	56.14	&	61.16	\\
CH$_3$CN	&	$J_{K,F}=6_{1,7}-5_{1,6}$	&	110.381404	&	44.55	&	36.40	&	33.69	&&	20.01	&	--	&	50.52	\\
C$^{17}$O	&	$J_F=1_{7/2}-0_{5/2}$	&	112.358988	&	29.32	&	31.99	&	45.33	&&	15.93	&	--	&	33.45	\\
CN	&	$N_{J,F}=1_{1/2,3/2}-0_{1/2,1/2}$	&	113.170528	&	51.73	&	49.08	&	9.35	&&	74.18	&	78.30	&	78.62	\\
CN	&	$N_{J,F}=1_{3/2,5/2}-0_{1/2,3/2}$	&	113.490982	&	73.92	&	23.73	&	39.58	&&	53.15	&	47.99	&	70.70	\\
CO	&	$J=1-0$	&	115.271202	&	52.54	&	34.34	&	29.13	&&	58.03	&	66.88	&	73.84
\enddata
\end{deluxetable}

\newpage

\begin{deluxetable}{rccccccccc}
\tablecolumns{10}
\tablewidth{0pc}
\tablecaption{Peak temperatures and 1$\sigma$ rms noise levels of the identified spectral lines}
\tablehead{
\colhead{} &  \colhead{}	& \colhead{Rest Freq.}	&	\multicolumn{3}{c}{$T_{\rm peak}$ (K)} &\colhead{} & \multicolumn{3}{c}{$\Delta T_{\rm MB}$ (K)} \\
\cline{4-6} \cline{8-10} 																	
\colhead{Species}	&	\colhead{Transition}	&	\colhead{(GHz)}	&	\colhead{NE}	&	\colhead{SGA}	&	\colhead{SW}	&\colhead{}&	\colhead{NE}	&	\colhead{SGA}	& \colhead{SW}	
}
\startdata															
HNO	&	$1_{0,1}-0_{0,0}$	&	81.477490	&	0.089	&	0.063	&	0.082	&&	0.023	&	0.022	&	0.023	\\
HCCCN	&	$J=9-8$	&	81.881463	&	0.320	&	0.252	&	0.328	&&	0.021	&	0.019	&	0.020	\\
c-C$_{3}$H$_{2}$	&	$2_{0,2}-1_{1,1}$	&	82.093555	&	0.061	&	0.048	&	0.048	&&	0.022	&	0.025	&	0.023	\\
CH$_3$OH	&	$5_{-1,5}-4_{0,4}$E	&	84.521206	&	0.263	&	0.208	&	0.196	&&	0.013	&	0.012	&	0.013	\\
OCS	&	$J=7-6$	&	85.139104	&	0.077	&	0.056	&	0.052	&&	0.019	&	0.019	&	0.020	\\
c-C$_3$H$_2$	&	$2_{1,2}-1_{0,1}$	&	85.338906	&	0.208	&	0.192	&	0.173	&&	0.019	&	0.021	&	0.018	\\
H	&	$42\alpha$	&	85.688180	&	0.122	&	0.091	&	0.158	&&	0.025	&	0.027	&	0.025	\\
H$^{13}$CN	&	$J=1-0$	&	86.340176	&	0.679	&	0.484	&	0.594	&&	0.022	&	0.023	&	0.021	\\
H$^{13}$CO$^{+}$	&	$J=1-0$	&	86.754288	&	0.194	&	0.123	&	0.197	&&	0.016	&	0.040	&	0.015	\\
SiO	&	$J=2-1$	&	86.846995	&	0.271	&	0.204	&	0.371	&&	0.022	&	0.018	&	0.018	\\
HN$^{13}$C	&	$J=1-0$	&	87.090859	&	0.128	&	0.140	&	0.111	&&	0.021	&	0.021	&	0.021	\\
CCH	&	$N_{J,F}=1_{3/2,2}-0_{1/2,1}$	&	87.316925	&	0.720	&	0.608	&	0.631	&&	0.022	&	0.024	&	0.021	\\
CCH	&	$N_{J.F}=1_{1/2,1}-0_{1/2,1}$	&	87.402004	&	0.424	&	0.361	&	0.344	&&	0.013	&	0.022	&	0.018	\\
HNCO	&	$4_{0,4}-3_{0,3}$	&	87.925238	&	0.233	&	0.275	&	0.307	&&	0.025	&	0.028	&	0.025	\\
HCN	&	$J=1-0$	&	88.631847	&	5.775	&	4.782	&	5.802	&&	0.033	&	0.052	&	0.030	\\
HCO$^+$	&	$J=1-0$	&	89.188526	&	4.189	&	2.637	&	4.267	&&	0.055	&	0.109	&	0.064	\\
HNC	&	$J=1-0$	&	90.663564	&	1.773	&	0.996	&	1.194	&&	0.014	&	0.014	&	0.016	\\
HCCCN	&	$J=10-9$	&	90.978993	&	0.277	&	0.185	&	0.228	&&	0.017	&	0.017	&	0.019	\\
CH$_3$CN	&	$J_{K,F}=5_{3,6}-4_{3,5}$	&	91.971310	&	0.144	&	0.092	&	0.133	&&	0.017	&	0.017	&	0.017	\\
H	&	$41\alpha$	&	92.034680	&	0.118	&	0.130	&	0.087	&&	0.020	&	0.022	&	0.016	\\
$^{13}$CS	&	$J=2-1$	&	92.494270	&	0.273	&	0.136	&	0.142	&&	0.017	&	0.019	&	0.016	\\
N$_2$H$^{+}$	&	$J_{F_1,F}=1_{1,2}-0_{1,2}$	&	93.171917	&	0.578	&	0.492	&	0.871	&&	0.028	&	0.036	&	0.031	\\
$^{13}$CH$_3$OH	&	$2_{0,2}-1_{0,1}$A++	&	94.407129	&	0.055	&	0.052	&	0.060	&&	0.019	&	0.022	&	0.019	\\
CH$_3$OH	&	$2_{1,2}-1_{1,1}$A++	&	95.914310	&	0.043	&	0.062	&	0.051	&&	0.013	&	0.011	&	0.012	\\
CH$_3$CHO	&	$5_{0,5}-4_{0,4}$A++	&	95.963465	&	0.068	&	0.064	&	0.052	&&	0.013	&	0.014	&	0.014	\\
C$^{34}$S	&	$J=2-1$	&	96.412950	&	0.289	&	0.115	&	0.162	&&	0.016	&	0.017	&	0.015	\\
CH$_3$OH	&	$2_{0,2}-1_{0,1}$E	&	96.744549	&	0.908	&	0.720	&	0.788	&&	0.014	&	0.013	&	0.014	\\
OCS	&	$J=8-7$	&	97.301209	&	0.094	&	0.066	&	0.069	&&	0.016	&	0.016	&	0.015	\\
CH$_3$OH	&	$2_{1,1}-1_{1,0}$A$--$	&	97.582808	&	0.078	&	0.062	&	0.063	&&	0.017	&	0.017	&	0.015	\\
CS	&	$J=2-1$	&	97.980953	&	3.673	&	1.914	&	1.840	&&	0.020	&	0.019	&	0.020	\\
H	&	$40\alpha$	&	99.022750	&	0.118	&	0.170	&	0.091	&&	0.016	&	0.023	&	0.017	\\
SO	&	$N_J=2_3-1_2$	&	99.299905	&	0.404	&	0.255	&	0.319	&&	0.023	&	0.020	&	0.020	\\
HCCCN	&	$J=11-10$	&	100.076386	&	0.308	&	0.240	&	0.191	&&	0.022	&	0.020	&	0.023	\\
CH$_3$OH	&	$8_{-2,7}-8_{1,7}$E	&	101.469719	&	0.115	&	0.067	&	0.106	&&	0.026	&	0.024	&	0.025	\\
H$_2$CS	&	$3_{0,3}-2_{0,2}$	&	103.040548	&	0.090	&	0.029	&	0.061	&&	0.016	&	0.015	&	0.017	\\
H$_2$CS	&	$3_{1,2}-2_{1,1}$	&	104.617109	&	0.125	&	0.091	&	0.069	&&	0.019	&	0.021	&	0.019	\\
H	&	$39\alpha$	&	106.737250	&	0.164	&	0.217	&	0.127	&&	0.023	&	0.034	&	0.025	\\
$^{13}$CN	&	$N_{J,F}=1_{1/2,1}-0_{1/2,2}$	&	108.658948	&	0.155	&	0.088	&	0.141	&&	0.023	&	0.023	&	0.024	\\
$^{13}$CN	&	$N_{J,F}=1_{3/2,2}-0_{1/2,1}$	&	108.782374	&	0.118	&	0.103	&	0.114	&&	0.021	&	0.021	&	--\tablenotemark{a}	\\
CH$_3$OH	&	$0_{0,0}-1_{-1,1}$E	&	108.893929	&	0.143	&	0.107	&	0.132	&&	0.019	&	0.025	&	0.023	\\
HCCCN	&	$J=12-11$	&	109.173638	&	0.340	&	0.217	&	0.220	&&	0.022	&	0.027	&	0.025	\\
OCS	&	$J=9-8$	&	109.463063	&	0.124	&	0.088	&	0.130	&&	0.021	&	0.022	&	0.023	\\
C$^{18}$O	&	$J=1-0$	&	109.782176	&	0.670	&	0.359	&	0.479	&&	0.021	&	0.019	&	0.024	\\
HNCO	&	$5_{0,5}-4_{0,4}$	&	109.905753	&	0.291	&	0.308	&	0.382	&&	0.020	&	0.021	&	0.023	\\
$^{13}$CO	&	$J=1-0$	&	110.201354	&	6.017	&	4.448	&	4.649	&&	0.019	&	0.020	&	0.020	\\
CH$_3$CN	&	$J_{K,F}=6_{1,7}-5_{1,6}$	&	110.381404	&	0.179	&	0.127	&	0.134	&&	0.026	&	0.032	&	0.026	\\
C$^{17}$O	&	$J_F=1_{7/2}-0_{5/2}$	&	112.358988	&	0.180	&	0.123	&	0.165	&&	0.035	&	0.032	&	0.038	\\
CN	&	$N_{J,F}=1_{1/2,3/2}-0_{1/2,1/2}$	&	113.170528	&	1.625	&	0.817	&	1.348	&&	0.072	&	0.061	&	0.045	\\
CN	&	$N_{J,F}=1_{3/2,5/2}-0_{1/2,3/2}$	&	113.490982	&	2.883	&	2.300	&	2.319	&&	0.071	&	0.074	&	0.073	\\
CO	&	$J=1-0$	&	115.271202	&	47.155	&	36.323	&	36.888	&&	0.141	&	0.285	&	0.234
\enddata
\tablenotetext{a}{{We could not calculate the rms noise here because of interference of adjacent lines.}}
\end{deluxetable}

\clearpage
\section{Discussion}
\subsection{Decomposition of Line Profiles}
All the identified lines contain multiple velocity components which originate from the +50 km s$^{-1}$ and +20 km s$^{-1}$ clouds, foreground/background spiral arms, and the CND.  The emission from the +50 km s$^{-1}$ and +20 km s$^{-1}$ clouds are dominant in many molecular lines toward NE and SW.  The +50 km s$^{-1}$ cloud appears in the NE spectra, while the +20 km s$^{-1}$ cloud appears in the SW spectra.
\par In order to investigate physical conditions and chemical composition of the CND, it is necessary to decompose the line profiles into their constituents.  However, it is not easy to perform the decomposition by spectral fitting, since each profile has complicated shape suffering from absorption and contamination of the disk gas.  
Hence we define velocity ranges which represent the CND and the GMCs, and calculate integrated intensities for each range at each position (Table 4).  
Figure 5 shows the defined velocity ranges in the HCN, CS, and CH$_3$OH line profiles.
Since the CND is rotating at a velocity of $\sim$110 km s$^{-1}$, the lines which trace the CND should be prominent in velocities around $\sim$+110 km s $^{-1}$ at NE and $\sim-$110 km s $^{-1}$ at SW.  This is well illustrated in the HCN and HCO$^+$ profiles.  The +50 km s$^{-1}$ and +20 km s$^{-1}$ clouds should be prominent in velocities around +50 km s$^{-1}$ at NE, and around +20 km s$^{-1}$ at SW, respectively.  These components are apparent in several optically thin lines such as the N$_2$H$^+$, CH$_3$OH and HCCCN lines.

\begin{deluxetable}{cccccccccccccc}
\tablecolumns{14}
\tablewidth{0pc}
\tablecaption{Velocity ranges for integration }
\tablehead{
&&\multicolumn{3}{c}{NE}&&\multicolumn{3}{c}{SGA}&&\multicolumn{3}{c}{SW}&\\
\cline{3-5} \cline{7-9} \cline{11-13}
&&total&GMC&CND&&total&GMC&CND&&total&GMC&CND&
}	
\startdata	
$V_{\rm min}$ (km s$^{-1}$)&&--150&30&120&&--200&20&70&&--200&10&--120&\\
$V_{\rm max}$ (km s$^{-1}$)&&200&60&90&&200&50&100&&150&40&--90&\\
\tableline
Symbol\tablenotemark{a}&&$I^{\rm NE}_{\rm total}$&$I^{\rm NE}_{\rm GMC}$&$I^{\rm NE}_{\rm CND}$&&$I^{\rm SGA}_{\rm total}$&$I^{\rm SGA}_{\rm GMC}$&$I^{\rm SGA}_{\rm CND}$&&$I^{\rm SW}_{\rm total}$&$I^{\rm SW}_{\rm GMC}$&$I^{\rm SW}_{\rm CND}$&\\
\enddata
\tablenotetext{a}{Symbol denotes the intensity integrated over $V_{\rm min}$ to $V_{\rm max}$.}
\end{deluxetable}

\begin{figure}[h]
\begin{center}
\epsscale{1}
\includegraphics{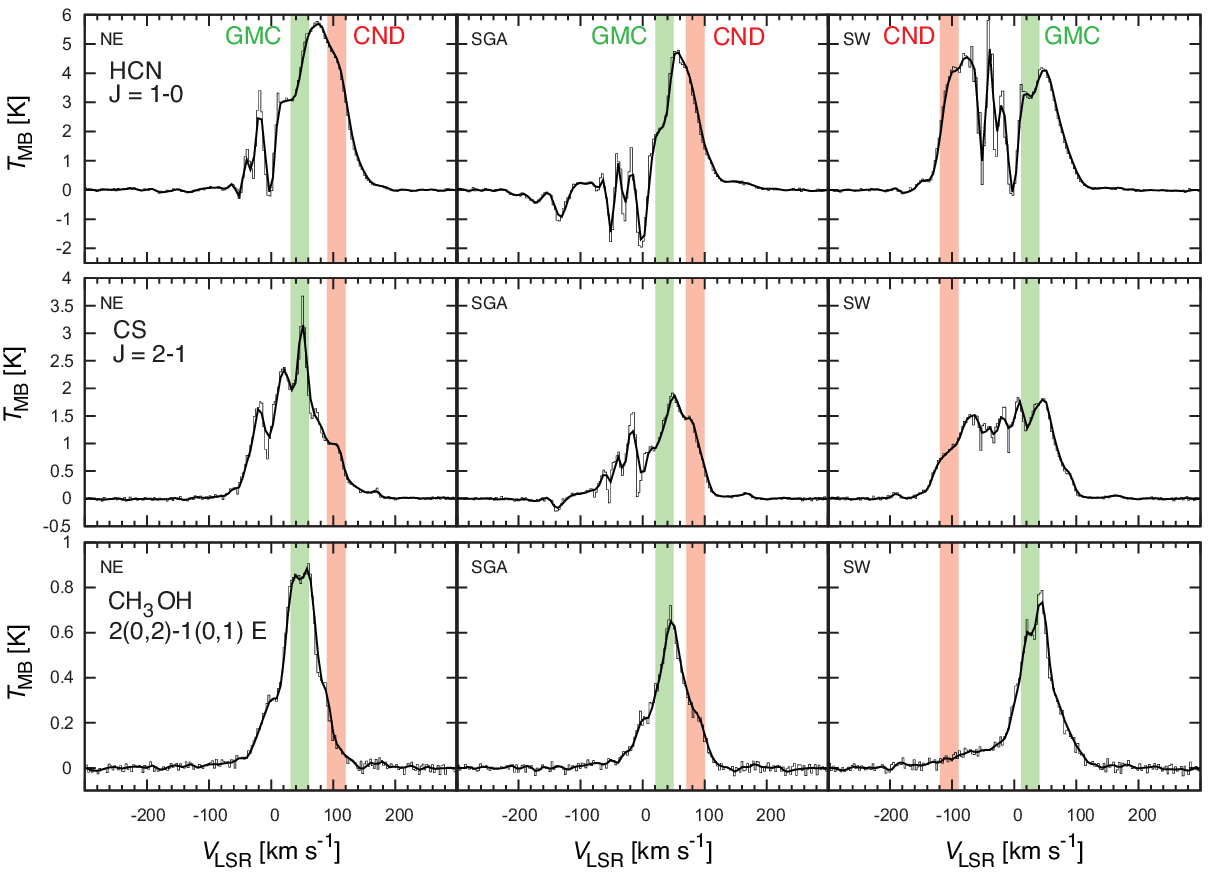}
\caption{Line profiles of the HCN, CS, and C$_3$OH lines toward NE, SW, and SGA.  The red and green bands show the velocity ranges which represent the CND and GMCs, respectively.  The HCN, CS, and CH$_3$OH lines are typical of the CND-, HBD-, and GMC-types, respectively (see \S4.3).  }
\end{center}
\end{figure}

\subsection{Line Intensities}　
We list the velocity-integrated line intensities (Table 5--7) calculated for the velocity ranges listed in Table 4.  
Figure 6 shows a correlation plot between the integrated intensities at NE and SW for the CND velocity range ($I^{\rm NE}_{\rm CND}$ and $I^{\rm SW}_{\rm CND}$, respectively).  The good correlation may support our choice of the velocity ranges which represent the same entity, the CND.   
We hence define that $I_{\rm CND} \equiv I^{\rm NE}_{\rm CND} + I^{\rm SW}_{\rm CND}$.  
Figure 7 shows a correlation plot between the integrated intensities at NE and SW for the GMC velocity range ($I^{\rm NE}_{\rm GMC}$ and $I^{\rm SW}_{\rm GMC}$, respectively).
{{The line intensities from the GMCs in NE are generally more intense than those in SW.  In other words, the +50 km s$^{-1}$ cloud is brighter than +20 km s$^{-1}$ cloud in most of the detected lines.}}
This plot also shows a tight correlation, although $I^{\rm NE}_{\rm GMC}$ is generally larger than $I^{\rm SW}_{\rm GMC}$. 
This fact indicates that the +50 km s$^{-1}$ and +20 km s$^{-1}$ clouds have roughly similar chemical composition and physical conditions.
Thus, we also define that $I_{\rm GMC} \equiv I^{\rm NE}_{\rm GMC} + I^{\rm SW}_{\rm GMC}$.  We list $I_{\rm CND}$ and $I_{\rm GMC}$ in Table 8, which values are most likely to represent the typical intensities from the CND and the GMCs.

\begin{deluxetable}{rccccccccc}
\tablecolumns{10}
\tablewidth{0pc}
\tablecaption{Velocity-integrated intensities of the identified lines toward NE }
\tablehead{
\colhead{}	&	\colhead{}	&	\colhead{Rest Freq.} &	\colhead{$I^{\rm NE}_{\rm total}$}	&	\colhead{$I^{\rm NE}_{\rm GMC}$}		&	\colhead{$I^{\rm NE}_{\rm CND}$}\\
\colhead{Species}	&	\colhead{Transition}	&	\colhead{(GHz)}	&	\colhead{(K$\cdot$km s$^{-1}$)}			&	\colhead{(K$\cdot$km s$^{-1}$)	}	&	\colhead{(K$\cdot$km s$^{-1}$)}}	
\startdata																	
HNO	&	$1_{0,1}-0_{0,0}$	&	81.477490	&	4.03	$\pm$	0.81	&	1.24	$\pm$	0.24	&	0.12	$\pm$	0.24	\\
HCCCN	&	$J=9-8$	&	81.881463	&	18.35	$\pm$	0.75	&	7.00	$\pm$	0.22	&	1.15	$\pm$	0.22	\\
c-C$_{3}$H$_{2}$	&	$2_{0,2}-1_{1,1}$	&	82.093555	&	1.34	$\pm$	0.78	&	0.76	$\pm$	0.23	&	0.29	$\pm$	0.23	\\
CH$_3$OH	&	$5_{-1,5}-4_{0,4}$E	&	84.521206	&	13.15	$\pm$	0.47	&	5.56	$\pm$	0.14	&	0.58	$\pm$	0.14	\\
OCS	&	$J=7-6$	&	85.139104	&	4.18	$\pm$	0.68	&	1.17	$\pm$	0.20	&	--	\\
c-C$_3$H$_2$	&	$2_{1,2}-1_{0,1}$	&	85.338906	&	16.41	$\pm$	0.64	&	3.86	$\pm$	0.19	&	2.08	$\pm$	0.19	\\
H	&	$42\alpha$	&	85.688180	&	9.55	$\pm$	0.84	&	1.25	$\pm$	0.25	&	2.13	$\pm$	0.25	\\
H$^{13}$CN	&	$J=1-0$	&	86.340176	&	54.17	$\pm$	0.76	&	13.41	$\pm$	0.22	&	8.50	$\pm$	0.22	\\
H$^{13}$CO$^{+}$	&	$J=1-0$	&	86.754288	&	11.41	$\pm$	0.36	&	2.96	$\pm$	0.11	&	2.52	$\pm$	0.11	\\
SiO	&	$J=2-1$	&	86.846995	&	27.57	$\pm$	0.66	&	5.80	$\pm$	0.19	&	3.92	$\pm$	0.19	\\
HN$^{13}$C	&	$J=1-0$	&	87.090859	&	7.88	$\pm$	0.73	&	2.53	$\pm$	0.21	&	0.48	$\pm$	0.21	\\
CCH	&	$N_{J,F}=1_{3/2,2}-0_{1/2,1}$	&	87.316925	&	68.32	$\pm$	0.53	&	16.30	$\pm$	0.16	&	5.77	$\pm$	0.16	\\
CCH	&	$N_{J.F}=1_{1/2,1}-0_{1/2,1}$	&	87.402004	&	36.51	$\pm$	0.34	&	9.75	$\pm$	0.10	&	2.57	$\pm$	0.10	\\
HNCO	&	$4_{0,4}-3_{0,3}$	&	87.925238	&	14.29	$\pm$	0.85	&	5.73	$\pm$	0.25	&	0.43	$\pm$	0.25	\\
HCN	&	$J=1-0$	&	88.631847	&	545.85	$\pm$	1.15	&	106.92	$\pm$	0.34	&	120.44	$\pm$	0.34	\\
HCO$^+$	&	$J=1-0$	&	89.188526	&	366.98	$\pm$	1.88	&	56.86	$\pm$	0.55	&	93.21	$\pm$	0.55	\\
HNC	&	$J=1-0$	&	90.663564	&	143.29	$\pm$	0.47	&	39.48	$\pm$	0.14	&	23.03	$\pm$	0.14	\\
HCCCN	&	$J=10-9$	&	90.978993	&	12.71	$\pm$	0.58	&	5.65	$\pm$	0.17	&	0.83	$\pm$	0.17	\\
CH$_3$CN	&	$J_{K,F}=5_{3,6}-4_{3,5}$	&	91.971310	&	7.99	$\pm$	0.37	&	1.23	$\pm$	0.12	&	0.25	$\pm$	0.12	\\
H	&	$41\alpha$	&	92.034680	&	4.17	$\pm$	0.42	&	1.16	$\pm$	0.14	&	0.53	$\pm$	0.14	\\
$^{13}$CS	&	$J=2-1$	&	92.494270	&	17.00	$\pm$	0.57	&	4.70	$\pm$	0.17	&	1.24	$\pm$	0.17	\\
N$_2$H$^{+}$	&	$J_{F_1,F}=1_{1,2}-0_{1,2}$	&	93.171917	&	38.15	$\pm$	0.67	&	14.05	$\pm$	0.20	&	1.54	$\pm$	0.20	\\
$^{13}$CH$_3$OH	&	$2_{0,2}-1_{0,1}$A++	&	94.407129	&	2.45	$\pm$	0.65	&	1.09	$\pm$	0.19	&	0.13	$\pm$	0.19	\\
CH$_3$OH	&	$2_{1,2}-1_{1,1}$A++	&	95.914310	&	1.40	$\pm$	0.26	&	0.65	$\pm$	0.09	&	--	\\
CH$_3$CHO	&	$5_{0,5}-4_{0,4}$A++	&	95.963465	&	2.19	$\pm$	0.26	&	0.82	$\pm$	0.09	&	0.41	$\pm$	0.09	\\
C$^{34}$S	&	$J=2-1$	&	96.412950	&	13.55	$\pm$	0.50	&	4.69	$\pm$	0.15	&	0.81	$\pm$	0.15	\\
CH$_3$OH	&	$2_{0,2}-1_{0,1}$E	&	96.744549	&	63.34	$\pm$	0.45	&	23.02	$\pm$	0.13	&	3.72	$\pm$	0.13	\\
OCS	&	$J=8-7$	&	97.301209	&	5.19	$\pm$	0.51	&	1.53	$\pm$	0.15	&	0.57	$\pm$	0.15	\\
CH$_3$OH	&	$2_{1,1}-1_{1,0}$A$--$	&	97.582808	&	3.00	$\pm$	0.54	&	0.95	$\pm$	0.16	&	0.19	$\pm$	0.16	\\
CS	&	$J=2-1$	&	97.980953	&	252.75	$\pm$	0.66	&	69.36	$\pm$	0.19	&	24.95	$\pm$	0.19	\\
H	&	$40\alpha$	&	99.022750	&	9.91	$\pm$	0.51	&	1.57	$\pm$	0.15	&	1.91	$\pm$	0.15	\\
SO	&	$N_J=2_3-1_2$	&	99.299905	&	27.81	$\pm$	0.74	&	8.00	$\pm$	0.22	&	2.42	$\pm$	0.22	\\
HCCCN	&	$J=11-10$	&	100.076386	&	13.16	$\pm$	0.62	&	5.17	$\pm$	0.18	&	0.69	$\pm$	0.18	\\
CH$_3$OH	&	$8_{-2,7}-8_{1,7}$E	&	101.469719	&	4.17	$\pm$	0.82	&	0.77	$\pm$	0.24	&	--	\\
H$_2$CS	&	$3_{0,3}-2_{0,2}$	&	103.040548	&	--	&	0.72	$\pm$	0.15	&	--	\\
H$_2$CS	&	$3_{1,2}-2_{1,1}$	&	104.617109	&	7.52	$\pm$	0.61	&	2.08	$\pm$	0.18	&	0.58	$\pm$	0.18	\\
H	&	$39\alpha$	&	106.737250	&	11.80	$\pm$	0.72	&	1.88	$\pm$	0.21	&	1.39	$\pm$	0.21	\\
$^{13}$CN	&	$N_{J,F}=1_{1/2,1}-0_{1/2,2}$	&	108.658948	&	11.95	$\pm$	0.54	&	1.93	$\pm$	0.16	&	2.11	$\pm$	0.16	\\
$^{13}$CN	&	$N_{J,F}=1_{3/2,2}-0_{1/2,1}$	&	108.782374	&	9.85	$\pm$	0.51	&	2.23	$\pm$	0.15	&	0.96	$\pm$	0.15	\\
CH$_3$OH	&	$0_{0,0}-1_{-1,1}$E	&	108.893929	&	2.67	$\pm$	0.42	&	2.48	$\pm$	0.12	&	-0.40	$\pm$	0.12	\\
HCCCN	&	$J=12-11$	&	109.173638	&	13.42	$\pm$	0.29	&	6.44	$\pm$	0.08	&	0.56	$\pm$	0.08	\\
OCS	&	$J=9-8$	&	109.463063	&	6.02	$\pm$	0.64	&	1.86	$\pm$	0.19	&	0.09	$\pm$	0.19	\\
C$^{18}$O	&	$J=1-0$	&	109.782176	&	32.21	$\pm$	0.54	&	12.36	$\pm$	0.16	&	0.60	$\pm$	0.16	\\
HNCO	&	$5_{0,5}-4_{0,4}$	&	109.905753	&	14.25	$\pm$	0.48	&	6.70	$\pm$	0.14	&	0.27	$\pm$	0.14	\\
$^{13}$CO	&	$J=1-0$	&	110.201354	&	425.35	$\pm$	0.58	&	149.80	$\pm$	0.17	&	13.10	$\pm$	0.17	\\
CH$_3$CN	&	$J_{K,F}=6_{1,7}-5_{1,6}$	&	110.381404	&	9.24	$\pm$	0.81	&	3.09	$\pm$	0.24	&	1.44	$\pm$	0.24	\\
C$^{17}$O	&	$J_F=1_{7/2}-0_{5/2}$	&	112.358988	&	9.49	$\pm$	1.08	&	3.62	$\pm$	0.32	&	0.09	$\pm$	0.32	\\
CN	&	$N_{J,F}=1_{1/2,3/2}-0_{1/2,1/2}$	&	113.170528	&	207.06	$\pm$	2.21	&	35.41	$\pm$	0.65	&	20.46	$\pm$	0.65	\\
CN	&	$N_{J,F}=1_{3/2,5/2}-0_{1/2,3/2}$	&	113.490982	&	274.16	$\pm$	2.15	&	47.48	$\pm$	0.63	&	47.22	$\pm$	0.63	\\
CO	&	$J=1-0$	&	115.271202	&	4203.28	$\pm$	4.26	&	1290.94	$\pm$	1.25	&	250.78	$\pm$	1.25	
\enddata
\end{deluxetable}

\begin{deluxetable}{rccccccccc}
\tablecolumns{10}
\tablewidth{0pc}
\tablecaption{Velocity-integrated intensities of the identified lines toward SGA }
\tablehead{
\colhead{}	&	\colhead{}	&	\colhead{Rest Freq.} &	\colhead{$I^{\rm SGA}_{\rm total}$}	&	\colhead{$I^{\rm SGA}_{\rm GMC}$}		&	\colhead{$I^{\rm SGA}_{\rm CND}$}\\
\colhead{Species}	&	\colhead{Transition}	&	\colhead{(GHz)}	&	\colhead{(K$\cdot$km s$^{-1}$)}			&	\colhead{(K$\cdot$km s$^{-1}$)	}	&	\colhead{(K$\cdot$km s$^{-1}$)}}	
\startdata																	
HNO	&	$1_{0,1}-0_{0,0}$	&	81.477490	&	5.21	$\pm$	0.59	&	0.71	$\pm$	0.23	&	0.26	$\pm$	0.23	\\
HCCCN	&	$J=9-8$	&	81.881463	&	14.89	$\pm$	0.60	&	5.23	$\pm$	0.20	&	1.94	$\pm$	0.20	\\
c-C$_{3}$H$_{2}$	&	$2_{0,2}-1_{1,1}$	&	82.093555	&	1.69	$\pm$	0.70	&	0.29	$\pm$	0.27	&	0.38	$\pm$	0.27	\\
CH$_3$OH	&	$5_{-1,5}-4_{0,4}$E	&	84.521206	&	10.64	$\pm$	0.38	&	3.90	$\pm$	0.12	&	1.36	$\pm$	0.12	\\
OCS	&	$J=7-6$	&	85.139104	&	4.46	$\pm$	0.57	&	0.73	$\pm$	0.20	&	0.53	$\pm$	0.20	\\
c-C$_3$H$_2$	&	$2_{1,2}-1_{0,1}$	&	85.338906	&	14.98	$\pm$	0.62	&	2.74	$\pm$	0.20	&	1.49	$\pm$	0.20	\\
H	&	$42\alpha$	&	85.688180	&	7.06	$\pm$	0.79	&	0.75	$\pm$	0.26	&	0.07	$\pm$	0.26	\\
H$^{13}$CN	&	$J=1-0$	&	86.340176	&	36.08	$\pm$	0.73	&	7.93	$\pm$	0.23	&	6.67	$\pm$	0.23	\\
H$^{13}$CO$^{+}$	&	$J=1-0$	&	86.754288	&	7.28	$\pm$	0.69	&	0.75	$\pm$	0.27	&	1.56	$\pm$	0.27	\\
SiO	&	$J=2-1$	&	86.846995	&	19.64	$\pm$	0.43	&	3.80	$\pm$	0.14	&	3.54	$\pm$	0.14	\\
HN$^{13}$C	&	$J=1-0$	&	87.090859	&	9.62	$\pm$	0.67	&	2.21	$\pm$	0.22	&	1.21	$\pm$	0.22	\\
CCH	&	$N_{J,F}=1_{3/2,2}-0_{1/2,1}$	&	87.316925	&	38.95	$\pm$	0.50	&	11.59	$\pm$	0.17	&	5.80	$\pm$	0.17	\\
CCH	&	$N_{J.F}=1_{1/2,1}-0_{1/2,1}$	&	87.402004	&	20.10	$\pm$	0.43	&	6.89	$\pm$	0.16	&	2.79	$\pm$	0.16	\\
HNCO	&	$4_{0,4}-3_{0,3}$	&	87.925238	&	20.67	$\pm$	0.91	&	5.69	$\pm$	0.28	&	2.15	$\pm$	0.28	\\
HCN	&	$J=1-0$	&	88.631847	&	307.64	$\pm$	1.55	&	71.94	$\pm$	0.52	&	81.32	$\pm$	0.52	\\
HCO$^+$	&	$J=1-0$	&	89.188526	&	149.07	$\pm$	2.83	&	11.43	$\pm$	1.09	&	53.28	$\pm$	1.09	\\
HNC	&	$J=1-0$	&	90.663564	&	59.06	$\pm$	0.38	&	19.78	$\pm$	0.14	&	9.76	$\pm$	0.14	\\
HCCCN	&	$J=10-9$	&	90.978993	&	9.77	$\pm$	0.50	&	3.61	$\pm$	0.17	&	1.16	$\pm$	0.17	\\
CH$_3$CN	&	$J_{K,F}=5_{3,6}-4_{3,5}$	&	91.971310	&	7.31	$\pm$	0.33	&	1.17	$\pm$	0.12	&	0.84	$\pm$	0.12	\\
H	&	$41\alpha$	&	92.034680	&	17.28	$\pm$	0.54	&	2.76	$\pm$	0.17	&	0.99	$\pm$	0.17	\\
$^{13}$CS	&	$J=2-1$	&	92.494270	&	18.65	$\pm$	0.59	&	2.80	$\pm$	0.19	&	2.77	$\pm$	0.19	\\
N$_2$H$^{+}$	&	$J_{F_1,F}=1_{1,2}-0_{1,2}$	&	93.171917	&	35.10	$\pm$	0.93	&	11.66	$\pm$	0.29	&	1.51	$\pm$	0.29	\\
$^{13}$CH$_3$OH	&	$2_{0,2}-1_{0,1}$A++	&	94.407129	&	3.05	$\pm$	0.54	&	0.73	$\pm$	0.22	&	0.10	$\pm$	0.22	\\
CH$_3$OH	&	$2_{1,2}-1_{1,1}$A++	&	95.914310	&	2.27	$\pm$	0.16	&	0.69	$\pm$	0.08	&	0.15	$\pm$	0.08	\\
CH$_3$CHO	&	$5_{0,5}-4_{0,4}$A++	&	95.963465	&	2.54	$\pm$	0.24	&	0.92	$\pm$	0.09	&	0.39	$\pm$	0.09	\\
C$^{34}$S	&	$J=2-1$	&	96.412950	&	8.31	$\pm$	0.45	&	1.85	$\pm$	0.15	&	1.12	$\pm$	0.15	\\
CH$_3$OH	&	$2_{0,2}-1_{0,1}$E	&	96.744549	&	41.49	$\pm$	0.40	&	14.13	$\pm$	0.13	&	6.85	$\pm$	0.13	\\
OCS	&	$J=8-7$	&	97.301209	&	4.48	$\pm$	0.42	&	0.83	$\pm$	0.15	&	0.67	$\pm$	0.15	\\
CH$_3$OH	&	$2_{1,1}-1_{1,0}$A$--$	&	97.582808	&	3.55	$\pm$	0.43	&	0.57	$\pm$	0.16	&	0.47	$\pm$	0.16	\\
CS	&	$J=2-1$	&	97.980953	&	169.51	$\pm$	0.60	&	37.88	$\pm$	0.18	&	31.44	$\pm$	0.18	\\
H	&	$40\alpha$	&	99.022750	&	28.90	$\pm$	0.76	&	2.95	$\pm$	0.21	&	1.41	$\pm$	0.21	\\
SO	&	$N_J=2_3-1_2$	&	99.299905	&	19.18	$\pm$	0.56	&	5.03	$\pm$	0.19	&	2.87	$\pm$	0.19	\\
HCCCN	&	$J=11-10$	&	100.076386	&	15.87	$\pm$	0.59	&	4.51	$\pm$	0.18	&	1.75	$\pm$	0.18	\\
CH$_3$OH	&	$8_{-2,7}-8_{1,7}$E	&	101.469719	&	4.16	$\pm$	0.63	&	0.76	$\pm$	0.23	&	0.11	$\pm$	0.23	\\
H$_2$CS	&	$3_{0,3}-2_{0,2}$	&	103.040548	&	2.16	$\pm$	0.34	&	0.18	$\pm$	0.14	&	0.08	$\pm$	0.14	\\
H$_2$CS	&	$3_{1,2}-2_{1,1}$	&	104.617109	&	6.32	$\pm$	0.58	&	1.52	$\pm$	0.20	&	0.70	$\pm$	0.20	\\
H	&	$39\alpha$	&	106.737250	&	29.72	$\pm$	1.11	&	3.76	$\pm$	0.31	&	2.00	$\pm$	0.31	\\
$^{13}$CN	&	$N_{J,F}=1_{1/2,1}-0_{1/2,2}$	&	108.658948	&	5.75	$\pm$	0.60	&	0.56	$\pm$	0.21	&	0.96	$\pm$	0.21	\\
$^{13}$CN	&	$N_{J,F}=1_{3/2,2}-0_{1/2,1}$	&	108.782374	&	5.60	$\pm$	0.37	&	0.98	$\pm$	0.14	&	0.97	$\pm$	0.14	\\
CH$_3$OH	&	$0_{0,0}-1_{-1,1}$E	&	108.893929	&	7.85	$\pm$	0.67	&	1.92	$\pm$	0.22	&	0.57	$\pm$	0.22	\\
HCCCN	&	$J=12-11$	&	109.173638	&	12.25	$\pm$	0.73	&	4.59	$\pm$	0.25	&	0.63	$\pm$	0.25	\\
OCS	&	$J=9-8$	&	109.463063	&	7.73	$\pm$	0.63	&	1.58	$\pm$	0.20	&	0.38	$\pm$	0.20	\\
C$^{18}$O	&	$J=1-0$	&	109.782176	&	20.23	$\pm$	0.46	&	6.92	$\pm$	0.15	&	1.69	$\pm$	0.15	\\
HNCO	&	$5_{0,5}-4_{0,4}$	&	109.905753	&	16.13	$\pm$	0.55	&	6.40	$\pm$	0.17	&	0.88	$\pm$	0.17	\\
$^{13}$CO	&	$J=1-0$	&	110.201354	&	294.15	$\pm$	0.67	&	91.72	$\pm$	0.18	&	29.57	$\pm$	0.18	\\
CH$_3$CN	&	$J_{K,F}=6_{1,7}-5_{1,6}$	&	110.381404	&	10.91	$\pm$	0.76	&	2.33	$\pm$	0.25	&	1.59	$\pm$	0.25	\\
C$^{17}$O	&	$J_F=1_{7/2}-0_{5/2}$	&	112.358988	&	8.82	$\pm$	0.80	&	1.58	$\pm$	0.28	&	0.68	$\pm$	0.28	\\
CN	&	$N_{J,F}=1_{1/2,3/2}-0_{1/2,1/2}$	&	113.170528	&	74.99	$\pm$	1.67	&	15.00	$\pm$	0.54	&	6.05	$\pm$	0.54	\\
CN	&	$N_{J,F}=1_{3/2,5/2}-0_{1/2,3/2}$	&	113.490982	&	113.44	$\pm$	1.82	&	29.51	$\pm$	0.66	&	19.16	$\pm$	0.66	\\
CO	&	$J=1-0$	&	115.271202	&	3046.59	$\pm$	9.15	&	891.18	$\pm$	2.51	&	300.58	$\pm$	2.51	
\enddata
\end{deluxetable}

\begin{deluxetable}{rccccccccc}
\tablecolumns{10}
\tablewidth{0pc}
\tablecaption{Velocity-integrated intensities of the identified lines toward SW }
\tablehead{
\colhead{}	&	\colhead{}	&	\colhead{Rest Freq.} &	\colhead{$I^{\rm SW}_{\rm total}$}	&	\colhead{$I^{\rm SW}_{\rm GMC}$}		&	\colhead{$I^{\rm SW}_{\rm CND}$}\\
\colhead{Species}	&	\colhead{Transition}	&	\colhead{(GHz)}	&	\colhead{(K$\cdot$km s$^{-1}$)}			&	\colhead{(K$\cdot$km s$^{-1}$)	}	&	\colhead{(K$\cdot$km s$^{-1}$)}}	
\startdata																																	
HNO	&	$1_{0,1}-0_{0,0}$	&	81.477490	&	1.89	$\pm$	0.83	&	1.17	$\pm$	0.24	&	0.11	$\pm$	0.24	\\
HCCCN	&	$J=9-8$	&	81.881463	&	17.84	$\pm$	0.71	&	5.98	$\pm$	0.21	&	1.16	$\pm$	0.21	\\
c-C$_{3}$H$_{2}$	&	$2_{0,2}-1_{1,1}$	&	82.093555	&	1.99	$\pm$	0.81	&	0.68	$\pm$	0.24	&	0.21	$\pm$	0.24	\\
CH$_3$OH	&	$5_{-1,5}-4_{0,4}$E	&	84.521206	&	10.53	$\pm$	0.45	&	3.81	$\pm$	0.13	&	0.45	$\pm$	0.13	\\
OCS	&	$J=7-6$	&	85.139104	&	1.86	$\pm$	0.72	&	0.75	$\pm$	0.21	&	0.49	$\pm$	0.21	\\
c-C$_3$H$_2$	&	$2_{1,2}-1_{0,1}$	&	85.338906	&	14.91	$\pm$	0.59	&	3.20	$\pm$	0.17	&	1.02	$\pm$	0.17	\\
H	&	$42\alpha$	&	85.688180	&	11.07	$\pm$	0.77	&	1.08	$\pm$	0.23	&	1.57	$\pm$	0.23	\\
H$^{13}$CN	&	$J=1-0$	&	86.340176	&	75.62	$\pm$	0.75	&	8.71	$\pm$	0.22	&	8.86	$\pm$	0.22	\\
H$^{13}$CO$^{+}$	&	$J=1-0$	&	86.754288	&	24.49	$\pm$	0.33	&	2.77	$\pm$	0.10	&	2.56	$\pm$	0.10	\\
SiO	&	$J=2-1$	&	86.846995	&	40.71	$\pm$	0.44	&	4.72	$\pm$	0.13	&	4.49	$\pm$	0.13	\\
HN$^{13}$C	&	$J=1-0$	&	87.090859	&	9.91	$\pm$	0.72	&	1.95	$\pm$	0.21	&	0.39	$\pm$	0.21	\\
CCH	&	$N_{J,F}=1_{3/2,2}-0_{1/2,1}$	&	87.316925	&	66.69	$\pm$	0.50	&	13.68	$\pm$	0.15	&	4.99	$\pm$	0.15	\\
CCH	&	$N_{J.F}=1_{1/2,1}-0_{1/2,1}$	&	87.402004	&	31.61	$\pm$	0.40	&	6.76	$\pm$	0.12	&	3.14	$\pm$	0.12	\\
HNCO	&	$4_{0,4}-3_{0,3}$	&	87.925238	&	15.24	$\pm$	0.88	&	5.39	$\pm$	0.26	&	0.36	$\pm$	0.26	\\
HCN	&	$J=1-0$	&	88.631847	&	601.92	$\pm$	1.02	&	90.43	$\pm$	0.30	&	94.43	$\pm$	0.30	\\
HCO$^+$	&	$J=1-0$	&	89.188526	&	316.63	$\pm$	2.20	&	56.29	$\pm$	0.65	&	39.47	$\pm$	0.65	\\
HNC	&	$J=1-0$	&	90.663564	&	153.53	$\pm$	0.53	&	27.81	$\pm$	0.16	&	17.37	$\pm$	0.16	\\
HCCCN	&	$J=10-9$	&	90.978993	&	10.91	$\pm$	0.64	&	3.95	$\pm$	0.19	&	0.84	$\pm$	0.19	\\
CH$_3$CN	&	$J_{K,F}=5_{3,6}-4_{3,5}$	&	91.971310	&	9.14	$\pm$	0.34	&	1.49	$\pm$	0.12	&	--	\\
H	&	$41\alpha$	&	92.034680	&	7.84	$\pm$	0.35	&	0.86	$\pm$	0.11	&	0.86	$\pm$	0.11	\\
$^{13}$CS	&	$J=2-1$	&	92.494270	&	14.72	$\pm$	0.53	&	2.67	$\pm$	0.16	&	1.16	$\pm$	0.16	\\
N$_2$H$^{+}$	&	$J_{F_1,F}=1_{1,2}-0_{1,2}$	&	93.171917	&	47.24	$\pm$	1.04	&	14.41	$\pm$	0.30	&	2.45	$\pm$	0.30	\\
$^{13}$CH$_3$OH	&	$2_{0,2}-1_{0,1}$A++	&	94.407129	&	--	&	0.86	$\pm$	0.19	&	--	\\
CH$_3$OH	&	$2_{1,2}-1_{1,1}$A++	&	95.914310	&	0.81	$\pm$	0.19	&	0.73	$\pm$	0.08	&	--	\\
CH$_3$CHO	&	$5_{0,5}-4_{0,4}$A++	&	95.963465	&	3.06	$\pm$	0.31	&	0.58	$\pm$	0.10	&	0.13	$\pm$	0.10	\\
C$^{34}$S	&	$J=2-1$	&	96.412950	&	16.09	$\pm$	0.48	&	2.76	$\pm$	0.14	&	1.62	$\pm$	0.14	\\
CH$_3$OH	&	$2_{0,2}-1_{0,1}$E	&	96.744549	&	50.68	$\pm$	0.48	&	15.93	$\pm$	0.14	&	1.19	$\pm$	0.14	\\
OCS	&	$J=8-7$	&	97.301209	&	3.05	$\pm$	0.48	&	1.04	$\pm$	0.14	&	--	\\
CH$_3$OH	&	$2_{1,1}-1_{1,0}$A$--$	&	97.582808	&	3.22	$\pm$	0.47	&	0.82	$\pm$	0.14	&	0.20	$\pm$	0.14	\\
CS	&	$J=2-1$	&	97.980953	&	259.57	$\pm$	0.66	&	41.77	$\pm$	0.19	&	23.79	$\pm$	0.19	\\
H	&	$40\alpha$	&	99.022750	&	6.47	$\pm$	0.51	&	0.43	$\pm$	0.15	&	1.06	$\pm$	0.15	\\
SO	&	$N_J=2_3-1_2$	&	99.299905	&	30.64	$\pm$	0.65	&	5.05	$\pm$	0.19	&	2.36	$\pm$	0.19	\\
HCCCN	&	$J=11-10$	&	100.076386	&	13.72	$\pm$	0.71	&	3.37	$\pm$	0.21	&	0.73	$\pm$	0.21	\\
CH$_3$OH	&	$8_{-2,7}-8_{1,7}$E	&	101.469719	&	3.77	$\pm$	0.79	&	1.19	$\pm$	0.23	&	0.42	$\pm$	0.23	\\
H$_2$CS	&	$3_{0,3}-2_{0,2}$	&	103.040548	&	2.94	$\pm$	0.55	&	0.96	$\pm$	0.16	&	0.18	$\pm$	0.16	\\
H$_2$CS	&	$3_{1,2}-2_{1,1}$	&	104.617109	&	2.02	$\pm$	0.59	&	1.00	$\pm$	0.17	&	0.18	$\pm$	0.17	\\
H	&	$39\alpha$	&	106.737250	&	9.18	$\pm$	0.77	&	1.38	$\pm$	0.23	&	2.05	$\pm$	0.23	\\
$^{13}$CN	&	$N_{J,F}=1_{1/2,1}-0_{1/2,2}$	&	108.658948	&	16.02	$\pm$	0.54	&	1.88	$\pm$	0.16	&	1.37	$\pm$	0.16	\\
$^{13}$CN	&	$N_{J,F}=1_{3/2,2}-0_{1/2,1}$	&	108.782374	&	13.87	$\pm$	0.69	&	1.92	$\pm$	0.22	&	1.67	$\pm$	0.22	\\
CH$_3$OH	&	$0_{0,0}-1_{-1,1}$E	&	108.893929	&	7.42	$\pm$	0.49	&	2.53	$\pm$	0.15	&	0.23	$\pm$	0.15	\\
HCCCN	&	$J=12-11$	&	109.173638	&	11.81	$\pm$	0.65	&	4.27	$\pm$	0.19	&	1.40	$\pm$	0.19	\\
OCS	&	$J=9-8$	&	109.463063	&	4.44	$\pm$	0.61	&	1.58	$\pm$	0.18	&	0.26	$\pm$	0.18	\\
C$^{18}$O	&	$J=1-0$	&	109.782176	&	29.53	$\pm$	0.64	&	8.41	$\pm$	0.19	&	0.67	$\pm$	0.19	\\
HNCO	&	$5_{0,5}-4_{0,4}$	&	109.905753	&	15.91	$\pm$	0.66	&	7.12	$\pm$	0.19	&	0.31	$\pm$	0.19	\\
$^{13}$CO	&	$J=1-0$	&	110.201354	&	336.54	$\pm$	0.62	&	103.36	$\pm$	0.18	&	8.86	$\pm$	0.18	\\
CH$_3$CN	&	$J_{K,F}=6_{1,7}-5_{1,6}$	&	110.381404	&	13.36	$\pm$	0.69	&	2.73	$\pm$	0.20	&	0.32	$\pm$	0.20	\\
C$^{17}$O	&	$J_F=1_{7/2}-0_{5/2}$	&	112.358988	&	14.24	$\pm$	1.15	&	2.22	$\pm$	0.34	&	1.31	$\pm$	0.34	\\
CN	&	$N_{J,F}=1_{1/2,3/2}-0_{1/2,1/2}$	&	113.170528	&	202.10	$\pm$	1.38	&	22.66	$\pm$	0.40	&	24.86	$\pm$	0.40	\\
CN	&	$N_{J,F}=1_{3/2,5/2}-0_{1/2,3/2}$	&	113.490982	&	252.42	$\pm$	2.21	&	37.26	$\pm$	0.65	&	42.22	$\pm$	0.65	\\
CO	&	$J=1-0$	&	115.271202	&	3576.56	$\pm$	7.07	&	971.28	$\pm$	2.07	&	249.15	$\pm$	2.07	
\enddata
\end{deluxetable}

\begin{deluxetable}{rccccccc}
\tablecolumns{8}
\tablewidth{0pc}
\tablecaption{Velocity-integrated intensities in the GMCs and CND, CND parameters, and spectral types}
\tablehead{
\colhead{}	&	\colhead{}	&	\colhead{Rest Freq.}	&   \colhead{$I_{\rm GMC}$}		&\colhead{$I_{\rm CND}$} &	\colhead{error (1$\sigma$)}	& \colhead{}& \\
\colhead{Species}	&	\colhead{Transition}	&	\colhead{(GHz)}	&	\colhead{(K$\cdot$km s$^{-1}$)}			&	\colhead{(K$\cdot$km s$^{-1}$)	}	&	\colhead{(K$\cdot$km s$^{-1}$)}& \colhead{$\xi_{\rm CND}$} & \colhead{Type}}	
\startdata										
HNO	&	$1_{0,1}-0_{0,0}$	&	81.477490	&	2.40	&	0.23	&	0.34	&	0.13	&	GMC	\\
HCCCN	&	$J=9-8$	&	81.881463	&	12.98	&	2.31	&	0.30	&	0.25	&	GMC	\\
c-C$_{3}$H$_{2}$	&	$2_{0,2}-1_{1,1}$	&	82.093555	&	1.43	&	0.51	&	0.33	&	0.50	&	HBD	\\
CH$_3$OH	&	$5_{-1,5}-4_{0,4}$E	&	84.521206	&	9.38	&	1.03	&	0.19	&	0.16	&	GMC	\\
OCS	&	$J=7-6$	&	85.139104	&	1.92	&	--	&	0.29	&	--	&	--	\\
c-C$_3$H$_2$	&	$2_{1,2}-1_{0,1}$	&	85.338906	&	7.05	&	3.10	&	0.25	&	0.63	&	HBD	\\
H	&	$42\alpha$	&	85.688180	&	2.33	&	3.70	&	0.33	&	2.24	&	CND	\\
H$^{13}$CN	&	$J=1-0$	&	86.340176	&	22.11	&	17.36	&	0.31	&	1.20	&	CND	\\
H$^{13}$CO$^{+}$	&	$J=1-0$	&	86.754288	&	5.73	&	5.08	&	0.15	&	1.26	&	CND	\\
SiO	&	$J=2-1$	&	86.846995	&	10.52	&	8.41	&	0.23	&	1.17	&	CND	\\
HN$^{13}$C	&	$J=1-0$	&	87.090859	&	4.47	&	0.87	&	0.30	&	0.28	&	GMC	\\
CCH	&	$N_{J,F}=1_{3/2,2}-0_{1/2,1}$	&	87.316925	&	29.98	&	10.75	&	0.22	&	0.51	&	HBD	\\
CCH	&	$N_{J.F}=1_{1/2,1}-0_{1/2,1}$	&	87.402004	&	16.51	&	5.70	&	0.16	&	0.53	&	HBD	\\
HNCO	&	$4_{0,4}-3_{0,3}$	&	87.925238	&	11.11	&	0.79	&	0.36	&	0.10	&	GMC	\\
HCN	&	$J=1-0$	&	88.631847	&	197.35	&	214.87	&	0.45	&	1.54	&	CND	\\
HCO$^+$	&	$J=1-0$	&	89.188526	&	113.16	&	132.68	&	0.85	&	1.78	&	CND	\\
HNC	&	$J=1-0$	&	90.663564	&	67.29	&	40.40	&	0.21	&	0.85	&	HBD	\\
HCCCN	&	$J=10-9$	&	90.978993	&	9.60	&	1.67	&	0.25	&	0.26	&	GMC	\\
CH$_3$CN	&	$J_{K,F}=5_{3,6}-4_{3,5}$	&	91.971310	&	2.73	&	--	&	0.17	&	--	&	--	\\
H	&	$41\alpha$	&	92.034680	&	2.01	&	1.39	&	0.18	&	1.10	&	CND	\\
$^{13}$CS	&	$J=2-1$	&	92.494270	&	7.37	&	2.41	&	0.23	&	0.51	&	HBD	\\
N$_2$H$^{+}$	&	$J_{F_1,F}=1_{1,2}-0_{1,2}$	&	93.171917	&	28.46	&	3.99	&	0.36	&	0.20	&	GMC	\\
$^{13}$CH$_3$OH	&	$2_{0,2}-1_{0,1}$A++	&	94.407129	&	1.95	&	--	&	0.27	&	--	&	--	\\
CH$_3$OH	&	$2_{1,2}-1_{1,1}$A++	&	95.914310	&	1.38	&	--	&	0.12	&	--	&	--	\\
CH$_3$CHO	&	$5_{0,5}-4_{0,4}$A++	&	95.963465	&	1.40	&	0.54	&	0.13	&	0.55	&	HBD	\\
C$^{34}$S	&	$J=2-1$	&	96.412950	&	7.45	&	2.44	&	0.20	&	0.61	&	HBD	\\
CH$_3$OH	&	$2_{0,2}-1_{0,1}$E	&	96.744549	&	38.95	&	4.91	&	0.19	&	0.18	&	GMC	\\
OCS	&	$J=8-7$	&	97.301209	&	2.57	&	0.56	&	0.21	&	0.37	&	--	\\
CH$_3$OH	&	$2_{1,1}-1_{1,0}$A$--$	&	97.582808	&	1.77	&	0.39	&	0.21	&	0.32	&	GMC	\\
CS	&	$J=2-1$	&	97.980953	&	111.13	&	48.75	&	0.27	&	0.67	&	HBD	\\
H	&	$40\alpha$	&	99.022750	&	2.01	&	2.97	&	0.21	&	2.72	&	CND	\\
SO	&	$N_J=2_3-1_2$	&	99.299905	&	13.05	&	4.78	&	0.29	&	0.56	&	HBD	\\
HCCCN	&	$J=11-10$	&	100.076386	&	8.55	&	1.42	&	0.28	&	0.25	&	GMC	\\
CH$_3$OH	&	$8_{-2,7}-8_{1,7}$E	&	101.469719	&	1.96	&	--	&	0.33	&	--	&	--	\\
H$_2$CS	&	$3_{0,3}-2_{0,2}$	&	103.040548	&	1.68	&	--	&	0.22	&	--	&	--	\\
H$_2$CS	&	$3_{1,2}-2_{1,1}$	&	104.617109	&	3.09	&	0.76	&	0.25	&	0.33	&	GMC	\\
H	&	$39\alpha$	&	106.737250	&	3.25	&	3.44	&	0.31	&	1.66	&	CND	\\
$^{13}$CN	&	$N_{J,F}=1_{1/2,1}-0_{1/2,2}$	&	108.658948	&	3.80	&	3.48	&	0.22	&	1.32	&	CND	\\
$^{13}$CN	&	$N_{J,F}=1_{3/2,2}-0_{1/2,1}$	&	108.782374	&	4.16	&	2.63	&	0.26	&	0.97	&	HBD	\\
CH$_3$OH	&	$0_{0,0}-1_{-1,1}$E	&	108.893929	&	5.00	&	--	&	0.19	&	--	&	--	\\
HCCCN	&	$J=12-11$	&	109.173638	&	10.71	&	1.96	&	0.21	&	0.34	&	GMC	\\
OCS	&	$J=9-8$	&	109.463063	&	3.44	&	0.35	&	0.26	&	0.17	&	GMC	\\
C$^{18}$O	&	$J=1-0$	&	109.782176	&	20.77	&	1.27	&	0.24	&	0.09	&	GMC	\\
HNCO	&	$5_{0,5}-4_{0,4}$	&	109.905753	&	13.82	&	0.58	&	0.24	&	0.06	&	GMC	\\
$^{13}$CO	&	$J=1-0$	&	110.201354	&	253.15	&	21.96	&	0.25	&	0.12	&	GMC	\\
CH$_3$CN	&	$J_{K,F}=6_{1,7}-5_{1,6}$	&	110.381404	&	5.83	&	1.76	&	0.31	&	0.48	&	GMC	\\
C$^{17}$O	&	$J_F=1_{7/2}-0_{5/2}$	&	112.358988	&	5.84	&	1.41	&	0.46	&	0.59	&	HBD	\\
CN	&	$N_{J,F}=1_{1/2,3/2}-0_{1/2,1/2}$	&	113.170528	&	58.08	&	45.31	&	0.76	&	1.24	&	CND	\\
CN	&	$N_{J,F}=1_{3/2,5/2}-0_{1/2,3/2}$	&	113.490982	&	84.74	&	89.44	&	0.90	&	1.51	&	CND	\\
CO	&	$J=1-0$	&	115.271202	&	2262.22	&	499.92	&	2.42	&	0.32	&	GMC	
\enddata
\end{deluxetable}
\clearpage

\begin{figure}[h]
\begin{center}
\epsscale{1}
\includegraphics{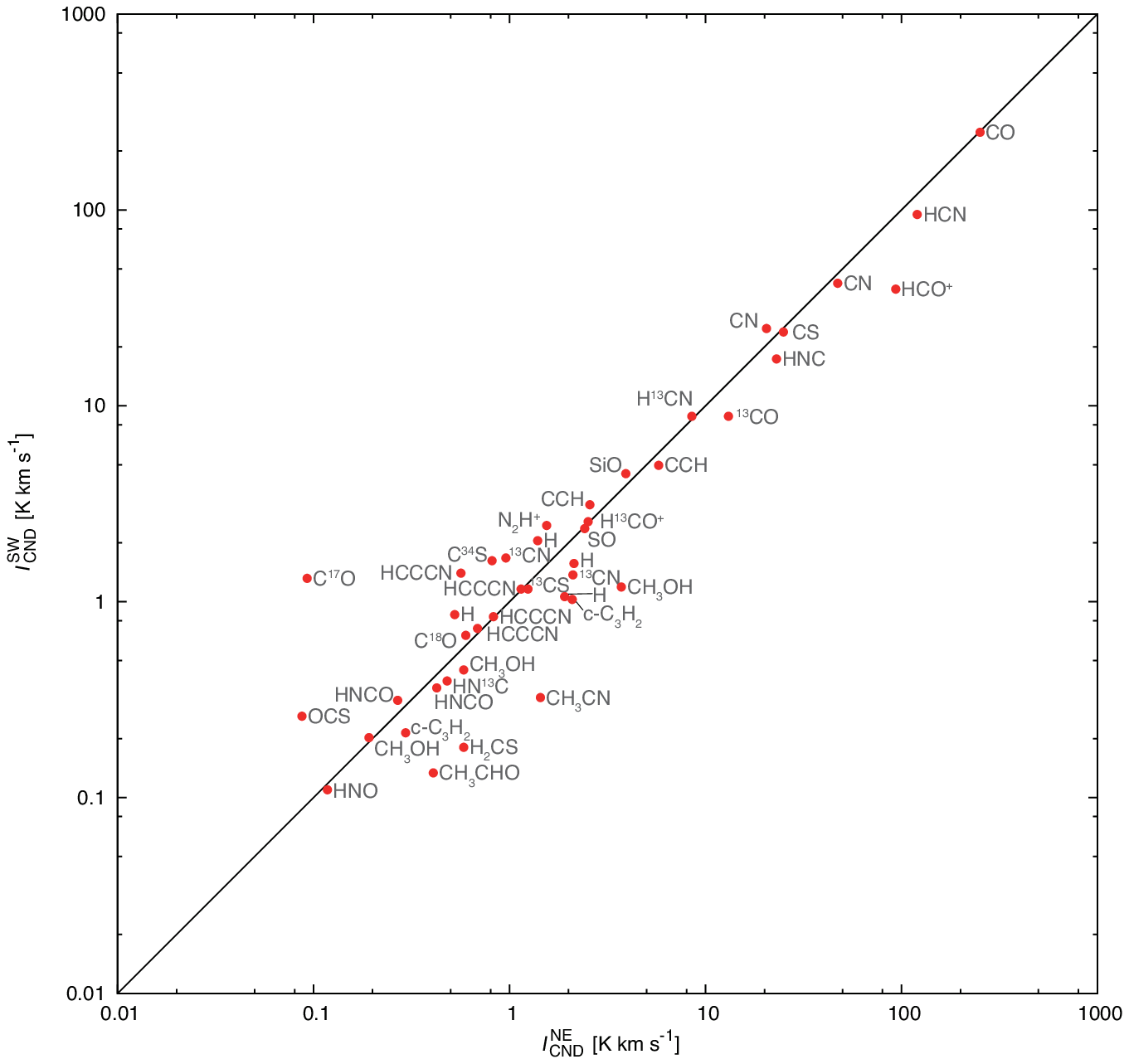}
\caption{{{Correlation plot between the integrated intensities $I^{\rm NE}_{\rm CND}$ and $I^{\rm SW}_{\rm CND}$ for each lines.}} A solid line shows where $I^{\rm NE}_{\rm CND}=I^{\rm SW}_{\rm CND}$.}
\end{center}
\end{figure}

\begin{figure}[h]
\begin{center}
\epsscale{1}
\includegraphics{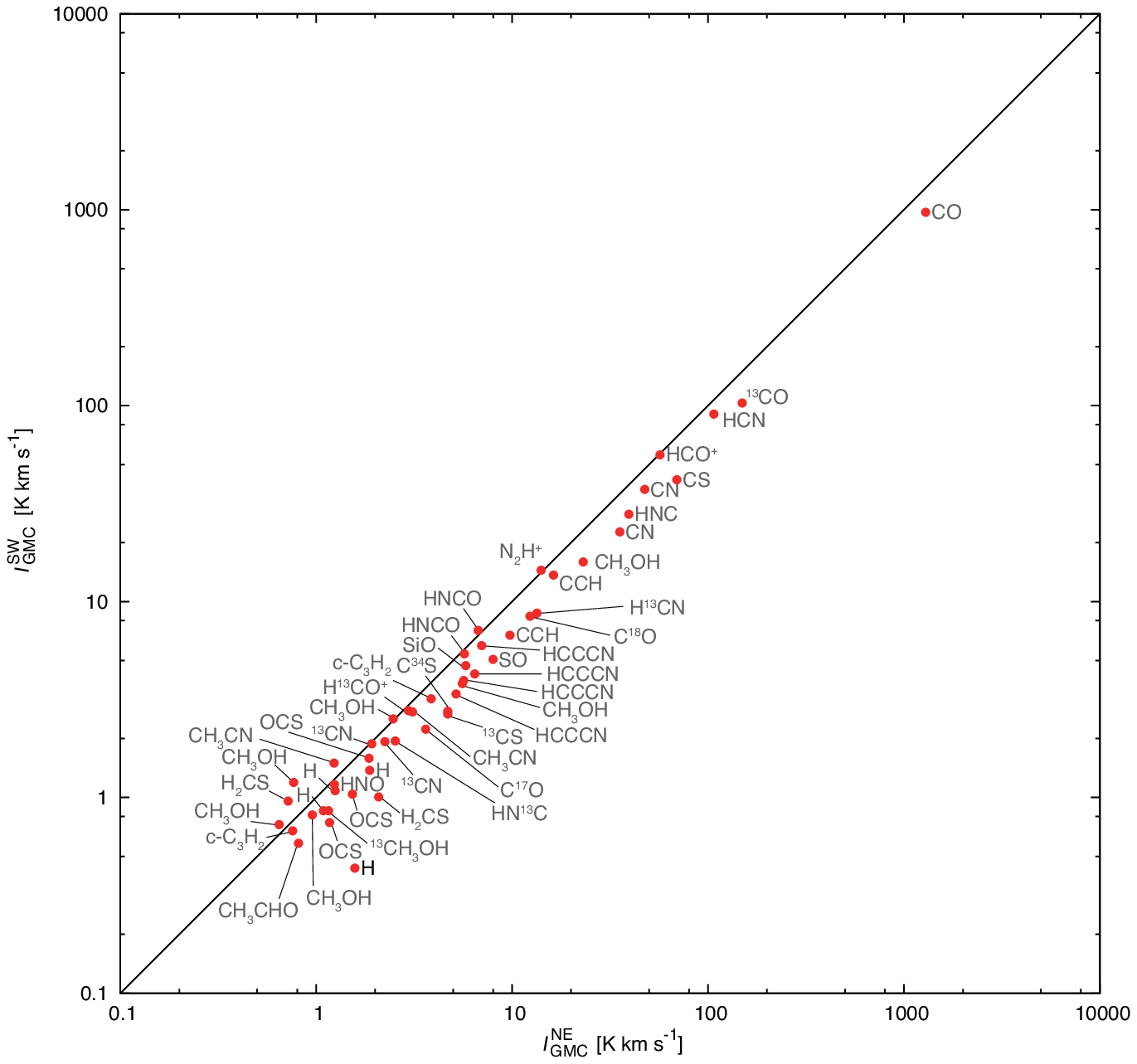}
\caption{{{Correlation plot between the integrated intensities $I^{\rm NE}_{\rm GMC}$ and $I^{\rm SW}_{\rm GMC}$ for each lines.}}  A solid line shows where $I^{\rm NE}_{\rm GMC}=I^{\rm SW}_{\rm GMC}$.}
\end{center}
\end{figure}

\subsection{Line Classification}
Figure 8 shows a plot of $I^{\rm NE}_{\rm CND}/I^{\rm NE}_{\rm GMC}$ versus $I^{\rm SW}_{\rm CND}/I^{\rm SW}_{\rm GMC}$ for each lines.  The lines in the upper right in the plot are enhanced toward the CND.
In order to search for prominent probes for the CND, we define the CND parameter: 
\begin{eqnarray}
\xi_{\rm CND} \equiv \sqrt{\left(\frac{I^{\rm NE}_{\rm CND}}{I^{\rm NE}_{\rm GMC}}\right)^2 + {\left(\frac{I^{\rm SW}_{\rm CND}}{I^{\rm SW}_{\rm GMC}}\right)}^2}.
\end{eqnarray}
  We classify the identified lines into three categories according to $\xi_{\rm CND}$;
\vspace{1mm}\\
\ \  (1) $\xi_{\rm CND} \geq 1$ : CND-type\\
\ \  (2) $0.5 \leq \xi_{\rm CND} < 1$ : Hybrid(HBD)-type\\
\ \  (3) $\xi_{\rm CND} < 0.5$ : GMC-type.\\
\vspace{1mm}\\
The CND- and GMC-types are the lines which mainly trace the CND and the GMCs, respectively. The HBD-type possesses the both characteristics of the CND/GMC-types. These types are also listed in Table 8.  For example, HCN, HCO$^+$ and SiO belong to the CND-type, HCCCN, N$_2$H$^+$ and CH$_3$OH belong to the GMC-type, and CS, HNC, CCH and SO belong to the HBD-type.  The lines of largish molecules tend to be the GMC-type.  The HBD-type lines arise from both of the CND and the GMCs, and thus these lines could probe physical connection between the CND and the GMCs.  
The line profiles of HCN, CS, and CH$_3$OH are shown in Figure 5 as representative examples of the CND-, HBD-, and GMC-types, respectively.

\begin{figure}[h]
\begin{center}
\epsscale{1}
\includegraphics{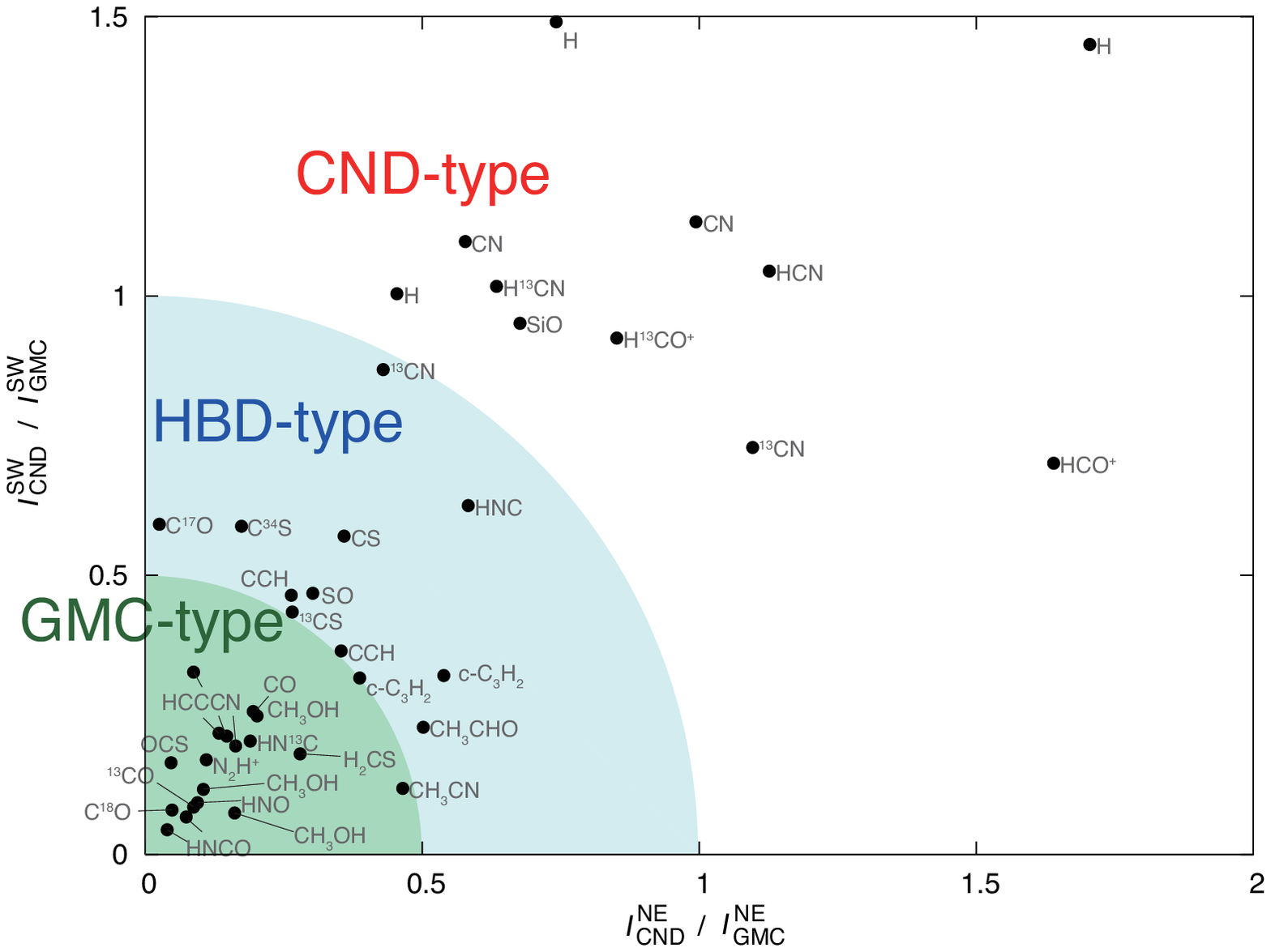}
\caption{Plot of $I^{\rm NE}_{\rm CND}/I^{\rm NE}_{\rm GMC}$ versus $I^{\rm SW}_{\rm CND}/I^{\rm SW}_{\rm GMC}$ for each lines. The green region satisfies $\xi_{\rm CND} < 0.5$ and species in this region are classified the GMC-type.  The turquoise region satisfies $0.5 \leq \xi_{\rm CND} < 1$ and species in this region are classified the HBD-type. Species out of arcs ($\xi_{\rm CND} \geq 1$) are classified the CND-type. }
\end{center}
\end{figure}

\subsection{Line Intensity Ratios}
Spectral line intensity ratios are used as indicators of physical conditions of molecular gas and chemical diagnosis, such as XDR/PDR/CRDR/shock-diagnosis.  We list several line intensity ratios which were calculated using $I_{\rm CND}$ and $I_{\rm GMC}$ in Table 9.
\par The CO/$^{13}$CO and HCN/H$^{13}$CN ratios can be used as opacity probes.
{{Both of the ratios}} are lower in the GMCs than in the CND, reflecting the higher optical depth of the CO and HCN lines in the GMCs. 
\par Ratios between dense gas and less dense gas probes are good indicators of gas density.  The HCN/CO, HCN/$^{13}$CO, H$^{13}$CN/$^{13}$CO, CS/CO, CS/$^{13}$CO, and $^{13}$CS/$^{13}$CO ratios can be used as density probes.
All these ratios are significantly higher in the CND than in the GMCs, likely indicating very high density in the CND.  In the Galactic disk (3.5 kpc $<R<$ 7 kpc) and the central 600 pc, the HCN/CO ratios are 0.026 $\pm$ 0.008 and 0.081$\pm$0.004, respectively (Helfer \& Blitz 1997).  
In the nuclear environment, the HCN/CO ratios are 0.0872 $\pm$ 0.0002 in the GMCs and 0.430 $\pm$ 0.002 in the CND.  The CS/CO ratios in the disk and the central 500 pc are 0.018 $\pm$ 0.008 and 0.027 $\pm$ 0.006, respectively (Helfer \& Blitz 1993, 1997).  The CS/CO ratios are 0.0491 $\pm$ 0.0001 in the nuclear GMCs and 0.0975 $\pm$ 0.0007 in the CND.  These trends in the ratios indicate a steep density gradient in molecular gas toward the nucleus.  

\par Several authors (e.g., Goldsmith et al. 1981; Churchwell et al. 1984; Schilke et al. 1992) predict that the HNC/HCN abundance ratio decreases with increasing temperature, so that the HNC/HCN ratio can be used as an indicator of gas temperature.  In nearby molecular clouds, the HNC/HCN ratio is known to range from $\sim$0.01 in hot cores (Schilke et al. 1992) up to $\sim$4 in dark clouds (Hirota et al. 1998).  In our data, the HNC/HCN ratios in the GMCs and the CND are 0.341 $\pm$ 0.001 and 0.188 $\pm$ 0.001, respectively.
 The HNC/HCN ratio, which is significantly lower in the CND than in the GMCs, may indicate higher temperature of the CND.  The HNC/H$^{13}$CN and HN$^{13}$C/H$^{13}$CN ratios show the same trends.  
\par The HCN/HCO$^+$ ratio exceeds unity ($>$1) in AGN, while the ratio is below unity ($<$1) in starburst galaxies (e.g. Kohno et al. 2004).  X-ray photons from AGN dissociate and ionize molecular gas, increasing the abundances of ions, radicals, and several molecular species.  These regions are called as XDRs (Lepp \& Dalgarno 1996; Maloney et al. 1996; Meijerink \& Spaans 2005; Meijerink et al. 2007).
In our data, the HCN/HCO$^+$ ratios are 1.74 $\pm$ 0.01 and 1.62 $\pm$ 0.01 in the GMCs and the CND, respectively.  They are comparable to each other, and significantly exceed unity.  Christopher et al. 2005 measured that the ratio in the CND was typically $\sim$ 2.5 with interferometric observations.  The ratios in the $l = 1.3^{\circ}$ complex and the Sgr B1 complex are $\sim$2.3 and $\sim$1.5, respectively (Tanaka et al. 2007, 2009).  Such high HCN/HCO$^+$ ratio is found over the whole of the central molecular zone (CMZ; e.g., Jones et al. 2012).
\par The SiO abundance is thought to be enhanced in shocked regions, and its intensity ratios to H$^{13}$CN and H$^{13}$CO$^+$ are often used as shock-probes.  
The SiO abundance in the CND seems not to be particularly enhanced, although the SiO line was categorized into the CND-type.  The SiO/H$^{13}$CO$^+$ ratios in the GMCs and the CND are 1.84 $\pm$ 0.06 and 1.66 $\pm$ 0.07, respectively.  These values are typical for the Sgr A molecular cloud complex (e.g., Tsuboi et al. 2011).

\begin{deluxetable}{ccc}
\tablecolumns{3}
\tablewidth{0pc}
\tablecaption{Line intensity ratios}
\tablehead{
\colhead{Ratio}  & \colhead{GMC} & \colhead{CND}
}	
\startdata		
CO/$^{13}$CO & 8.94 $\pm$ 0.01& 22.8 $\pm$ 0.3 \\
HCN/H$^{13}$CN & 8.9 $\pm$ 0.1 & 12.4 $\pm$ 0.2 \\
HCN/CO & 0.0872 $\pm$ 0.0002 & 0.430 $\pm$ 0.002 \\
HCN/$^{13}$CO & 0.780 $\pm$ 0.002 & 9.8 $\pm$ 0.1 \\
H$^{13}$CN/$^{13}$CO & 0.087 $\pm$ 0.001 & 0.79 $\pm$ 0.02 \\
CS/CO & 0.0491 $\pm$ 0.0001& 0.0975 $\pm$ 0.0007 \\
CS/$^{13}$CO & 0.439 $\pm$ 0.001 & 2.22 $\pm$ 0.03 \\
$^{13}$CS/$^{13}$CO & 0.0291 $\pm$ 0.0009 & 0.11 $\pm$ 0.01 \\
HNC/HCN & 0.341 $\pm$ 0.001& 0.188 $\pm$ 0.001 \\
HNC/H$^{13}$CN & 3.04 $\pm$ 0.04& 2.33 $\pm$ 0.04 \\
HN$^{13}$C/H$^{13}$CN & 0.20 $\pm$ 0.01 & 0.05 $\pm$ 0.02 \\
HCN/HCO$^+$& 1.74 $\pm$ 0.01& 1.62 $\pm$ 0.01 \\
H$^{13}$CN/H$^{13}$CO$^+$& 3.9 $\pm$ 0.1 & 3.4 $\pm$ 0.1 \\
SiO/H$^{13}$CN& 0.48 $\pm$ 0.01 & 0.48 $\pm$ 0.02 \\
SiO/H$^{13}$CO$^+$& 1.84 $\pm$ 0.06 & 1.66 $\pm$ 0.07
\enddata
\end{deluxetable}

\subsection{Spectra from Sgr A$^*$}
Suffering from severe absorption features against the intense continuum radiation from Sgr A$^*$, decomposition of line profiles toward SGA is highly difficult.  Based on the past literatures (e.g., G{\"u}sten et al. 1987; Kaifu et al. 1987; Oka et al. 2011), we define $V_{\rm LSR}=$+70 to +100 km s$^{-1}$ as the CND range, and $V_{\rm LSR}=$+20 to +50 km s$^{-1}$ as the GMC range (Figure 5).  
We calculated the velocity-integrated intensities of SGA ($I^{\rm SGA}_{\rm tot}$, $I^{\rm SGA}_{\rm GMC}$, and $I^{\rm SGA}_{\rm CND}$), using data with $T_{\rm MB} >0$ K in order to minimize the effect of the absorption by the foreground.
These intensities are listed in Table 6.  Figure 9 shows a correlation plot of the GMC contribution ($I^{\rm SGA}_{\rm GMC}/I^{\rm SGA}_{\rm tot}$) versus the CND contribution ($I^{\rm SGA}_{\rm CND}/I^{\rm SGA}_{\rm tot}$).  
The GMC-types prefer the top left of the plot, while the CND-types roughly prefer the bottom right.  
The HBD-types are distributed between the CND-types and GMC-types.
This may support the validity of our line classification and that of our definition of velocity ranges on the SGA spectra. 

\begin{figure}[h]
\begin{center}
\includegraphics{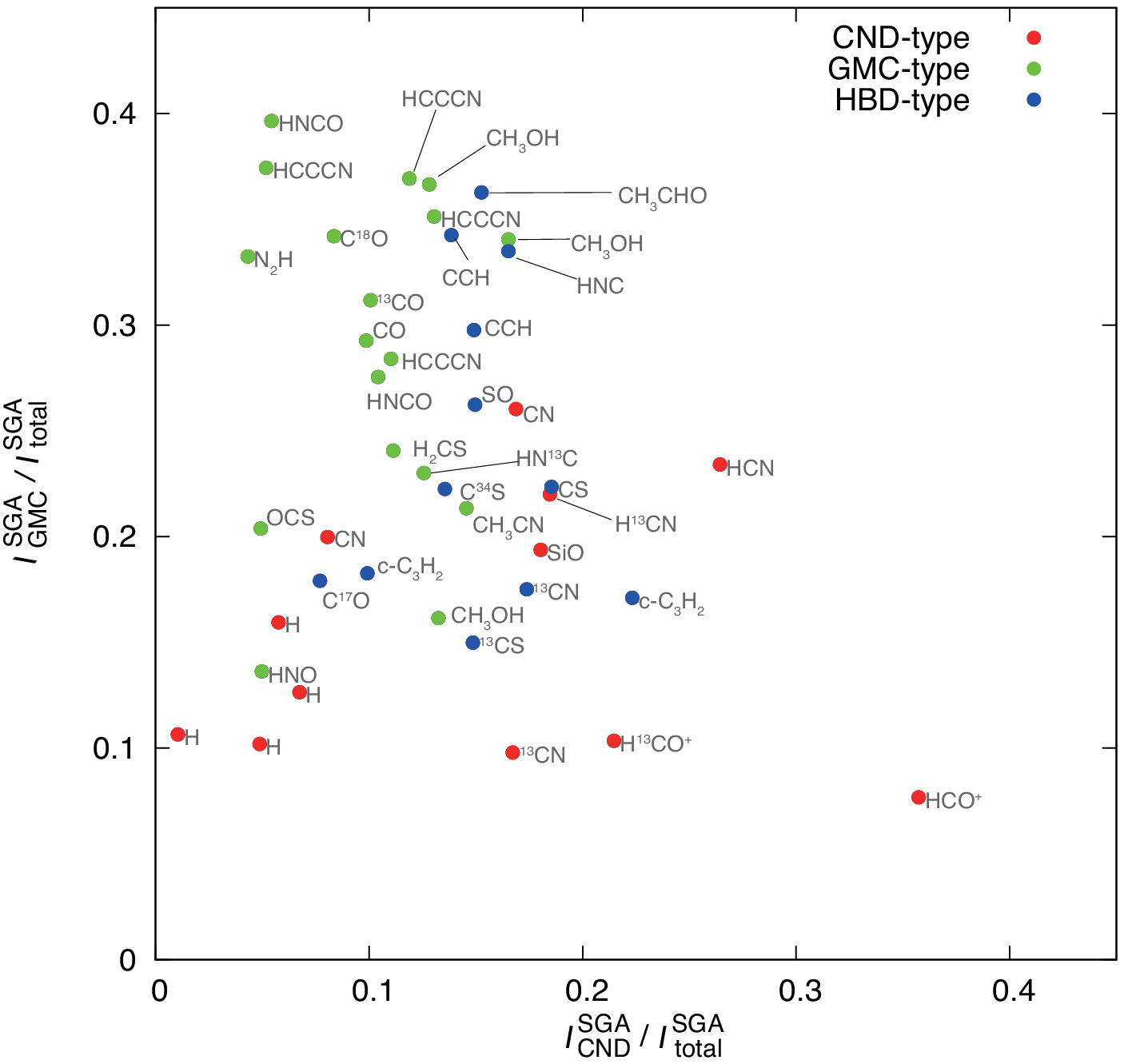}
\caption{Correlation plot between $I^{\rm SGA}_{\rm GMC}/I^{\rm SGA}_{\rm total}$ and $I^{\rm SGA}_{\rm CND}/I^{\rm SGA}_{\rm total}$ {\bf{for each lines}}.  Red, blue, and green circles represent the CND-, HBD-, and GMC-type lines, respectively.  
}
\end{center}
\end{figure}

\clearpage
\section{Summary}
We performed unbiased spectral line surveys at 3 mm band toward the Galactic CND and Sgr A$^*$ using the NRO 45 m radio telescope.  The target positions were two tangential points of the CND and the direction of Sgr A$^*$ (NE, SW, and SGA).  
With these surveys, we obtained three wide-band spectra which cover the frequency range from 81.3 GHz to 115.8 GHz, detecting 46 molecular lines from 30 species including 10 rare isotopomers and four hydrogen recombination lines.
The detected lines consist of multiple velocity components which arise from the CND, GMCs (+50 km s$^{-1}$ and +20 km s$^{-1}$ clouds), and the foreground spiral arms.  Many of the line profiles toward SGA severely suffer from absorption features.
\par We defined the specific velocity ranges which represent the CND and the GMCs toward each direction.
Based on the line intensities integrated over the defined velocity ranges, we classified the detected lines into three categories: the CND-/HBD-/GMC-types.  The rotational lines of HCN, H$^{13}$CN, HCO$^+$, H$^{13}$CO$^+$, SiO, CN, and $^{13}$CN, and hydrogen recombination lines belong to the CND-type. 
The detected lines include many diagnostic probes, i.e., opacity, density, temperature, XDR, and shock.  We presented the lists of the line intensities and intensity ratios, which must be useful to investigate the difference between nuclear environments of our Galaxy and of others.
Deep mapping observations in the CND-type lines
{{with a single dish}} would reveal the accurate distribution and kinematics of molecular gas in the vicinity of our Galactic nucleus.  {{We have already conducted such mapping observations with the NRO 45 m telescope.  These results and detailed analyses will be presented in forthcoming papers.}}
\acknowledgments
We are grateful to the NRO staff for their excellent support of the 45 m observations.
The Nobeyama Radio Observatory is a branch of the National Astronomical Observatory of Japan, National Institutes of Natural Sciences.
{{We thank the anonymous referee for his/her constructive comments which improve this paper.}}

\end{document}